\documentclass{aa}
\usepackage{soul}
\usepackage{graphicx} % Required for inserting images
\usepackage{txfonts}
\usepackage{natbib}
\usepackage{hyperref}
\usepackage{siunitx}
\usepackage{lineno}
\usepackage{xcolor}
\usepackage[english]{babel}
\usepackage{url}

\def\matteo#1{{#1}}

\begin{document}
\title{Born to be recycled: a comprehensive population synthesis of the Galactic millisecond pulsars}
\author{Mattéo Sautron\inst{1} \and Jérôme Pétri\inst{1} \and Dipanjan Mitra\inst{2,4} \and Adélie Dupuy{-}{-}Junet\inst{1,3} \and Marie-Eloïse Pietrin\inst{1,3}}
\institute{Université de Strasbourg, CNRS, Observatoire astronomique de Strasbourg, UMR 7550, 67000 Strasbourg, France. \and National Centre for Radio Astrophysics, Tata Institute for Fundamental Research, Post Bag 3, Ganeshkhind, Pune 411007, India. \and Université de Strasbourg, Télécom Physique Strasbourg, F-67400 Illkirch-Graffenstaden, France.
\and Janusz Gil Institute of Astronomy, University of Zielona G\'ora, ul. Szafrana 2, 65-516 Zielona G\'ora, Poland.}
\date{Received / Accepted}
\titlerunning{The Galactic recycled pulsars}
\authorrunning{Sautron et al.}

\abstract{
Millisecond pulsars (MSPs) are the oldest but fastest pulsars known to date, the first one, PSR B1937+21 was discovered in 1982. To explain how these pulsars could be formed, a new hypothesis was formulated: the recycling of pulsars, i.e the fact that a pulsar could accrete matter from a companion and been spun up.}
{In this paper, we developed a population synthesis algorithm for pulsars which belong to a binary, in order to check whether most of the observed recycled pulsars were formed via an accretion mechanism and derive statistics about their properties, that are difficult to obtain through observations. We also make predictions for future surveys.}
{Toward the presented objectives, we use the code Stellar EVolution for N-body (SEVN) to take into account all the binary processes and our own code to evolve each pulsar self-consistently by taking into account the secular evolution of a force-free magnetosphere, the magnetic field decay, gravitational braking and spatial evolution. Each pulsar is born in binary with a main sequence companion, and evolve to present time. The radio and $\gamma$-ray emission locations were modeled  by the polar cap geometry and striped wind model, respectively.}
{Our simulations reproduce well the population of radio and $\gamma$-ray pulsars observed in the selected surveys, as supported by Kolmogorov-Smirnov (KS) tests. We also found that there should be less than \matteo{$220$} unidentified pulsars in the Fourth Fermi-LAT catalogue of $\gamma$-ray sources (4FGL). Most of the pulsars accrete for typical durations of $10^{8-9}$~yr. Moreover the typical timescale of alignment between the rotation axis and the normal to the orbital plane (angle $\alpha$) remains usually much shorter. As a consequence, $\sim$80\% of the binary pulsars population has $\alpha$ close to 0°. Most of recycled pulsars possess masses around 1.8~$M_{\odot}$, some having masses up to 2.7~$M_{\odot}$, in agreement with recent observations of spiders pulsars which are massive. We estimate that $\sim190$ MSPs originally born in the Galactic spiral arms contribute to $\sim5$~\% of the GeV excess, reinforcing the fact that if the GeV excess is entirely of pulsar origin, these MSPs must have come from globular clusters that migrated inward to the Galactic center, or were born in the bulge. High values of the viewing angle $\zeta$ seem to be needed to be able to observe the recycled pulsars, and it also seems difficult to observe recycled pulsars with an aligned rotation axis and magnetic axis (i.e., $\chi \leq 10$°). We find that only a small fraction, approximately \matteo{$\sim 7.6\times10^{-3}$~\%}, of oxygen-neon white dwarfs (ONeWDs) in binary systems appear to contribute to the population of mildly recycled pulsars through accretion-induced collapse.  
}
{}

\maketitle
\nolinenumbers
\section{Introduction} \label{sec:intro}
Millisecond pulsars (MSPs) represent the oldest objects in the entire population of pulsars. Although they are at the very end of the stellar evolution, they are also the fastest spinning pulsars known (with periods $P < 30$~ms). In contrast to normal isolated pulsars and magnetars, MSPs are found in binary system, a fact that provides strong evidence for the rapid rotation. During the binary evolution stage they likely accreted matter from their companion, which increased their angular momentum and spun them up to high rotational frequencies \citep{acr+82,bv91,tlk12}. More broadly, these objects are referred to as `recycled pulsars", as they are thought to have initially undergone a significant spin-down before being spun up again. During the accretion phase, these sources are observed as X-rays binaries \citep{wv98,pw21}. Although different mechanisms (accretion process, magnetic field decay) are involved in recycling of pulsars, these  are not completely understood. Nevertheless the recycling scenario is widely accepted in the community today \citep{rs82,yl23,th23}. MSPs are also observed without a companion and their properties compared to recycled pulsars with a companion are not distinguishable, which support the idea that all recycled pulsars must have been related to binary evolution at some point \citep{sh82}. Moreover recycled pulsars appear in various forms. Some are isolated, as mentioned earlier \citep{bkh+82}, others have a companion that can be degenerate or not, and in some cases the MSP in the binary can be considered as a spider. These include black widows where the pulsar's very light companion ($M < 0.05 M_{\odot}$) is being destroyed by the pulsar wind and redbacks which have more massive companions ($M \sim$ $0.1-0.4 M_{\odot}$) that are ablated by the pulsar \citep{r13,bgv+25}. Furthermore, transitional MSPs are sources which are switching between X-ray and radio emitters \citep{pm22}. In eclipsing MSPs, the radio emission is eclipsed by the companion and could be due to ejected ionized material from the companion which is absorbing the radio waves from the MSP \citep{fst88}. Finally, MSPs with planets also exist \citep{wf92}. 

Until 2008, very few pulsars were detected in $\gamma$-rays (about 7 \citep{t08}), it was thank to the Fermi Gamma-Ray Space Telescope (Fermi) that this number could be significantly increased to about 300 today \citep{saa+23}. Only a few MSPs are detected in gamma-rays without a radio counterpart. Already in the 1990s, it was proposed that many MSPs might be $\gamma$-ray sources \citep{s90}, and Fermi has since led to a dramatic increase in the number of detected $\gamma$-ray MSPs, although the vast majority of known MSPs have been discovered through radio observations. Radio pulsar surveys, such as those conducted with the Parkes, Arecibo, Green Bank telescopes, Five-hundred-meter Aperture Spherical Telescope (FAST), MeerKAT, have identified over 400 MSPs to date \citep{mht+05,qys+20,bja+20}, many of which were later found to emit in $\gamma$-rays as well. Radio observations remain essential for accurately determining pulsar timing parameters, detecting binary companions, and characterizing their orbital dynamics. Furthermore, several MSPs are still only visible in the radio domain, highlighting the complementarity of multiwavelength studies.

A powerful tool for investigating neutron star evolution is population synthesis. Numerous studies have applied this method to different neutron star families, including the normal pulsar population \citep{fk06, gmv+14, spm+24, prg+25}, magnetars \citep{jt22, bhv+19, smy+25}, and MSPs \citep{khb+08, sgh07, ghf+18, tl25}. Through these studies, we can extract from simulations invaluable insights into the underlying population properties such as the birth spin period, magnetic field decay timescale, and birth rate, as well as the emission processes at play. In this approach, neutron stars are generated at birth and evolved to the present epoch, after which detection criteria are applied to determine whether they would be observable with current or future instruments.

We focus exclusively on mildly or fully recycled pulsars (a mildly recycled pulsar has accreted matter, but not enough to become fully recycled with $P < 30$~ms, thus it has $30\leq P \leq1000$~ms and a $\dot{P}<10^{-16}$), that are not in globular clusters (GCs). Although, about half of the known population of MSPs belong to a GC, studying pulsars that are in GCs is very different. Indeed, to study them accurately with simulations we would need to take many interactions into account. Since there are high stellar densities in these environments (much higher than the Galactic field), we would need to take into account the fact that there could be exchange of companions (at least it should happen more compared to the Galactic field), triple systems, there might be different distribution at birth for the eccentricity or the semi-major axis of the binary. For instance \citet{bef+24} used simulation-based inference (SBI, \citet{cbl+20}) to study MSPs population in GCs, and they focused on Terzan 5 by finding constraints on the luminosity function of its MSPs and the number (with uncertainties) of MSPs that Terzan 5 hosts.

The studies of \citet{sgh07,ghf+18} are the most complete studies about population synthesis of MSPs. The less recent study of \citet{sgh07} made predictions for Fermi and revealed that many MSPs would be detected by it as radio quiet $\gamma$-ray sources, which was indeed true \citep{aaa+13,sbc+19,saa+23}. In their paper they started the evolution of MSPs after they finished their spun-up episode, the magnetic field of MSPs was constant and they only considered MSPs from the Galactic disk, while the more recent study of \citet{ghf+18} improved the latter point by considering MSPs in the bulge too. They predicted that 11000 MSPs would be necessary in the Galactic bulge to explain the GeV excess detected in the Galactic center and their results are in agreement with those derived from Fermi detections. Although the results of \citet{ghf+18} are compelling, our approach, which self-consistently models the full binary evolution alongside the pulsar’s evolution and a detection model in $\gamma$-ray and radio to reproduce the recycled population, offers several advantages. It is so far the only study allowing to follow the full evolution to the MSP state, and even to produce mildly recycled pulsars, that takes into account the detectability in both radio and $\gamma$-ray wavelengths. This constitutes a valuable opportunity to investigate whether the accretion-driven formation channel is an effective pathway for producing MSPs, especially in comparison with alternative scenarios such as the collapse of white dwarfs into rapidly rotating neutron stars. Moreover, it enables the derivation of statistical properties of the recycled population, which can be otherwise difficult to access observationally. Finally, it allows us to make predictions for upcoming surveys such as Square Kilometre Array (SKA \footnote{\url{https://www.skao.int/en/science-users/118/ska-telescope-specifications}}).% and Cherenkov Telescope Array (CTA \footnote{\url{https://www.cta-observatory.org/about/how-ctao-works/}}).  

This paper explores many aspects of binary evolution leading to recycled pulsars. It is organised as follow: in Sec.~\ref{sec:model}, we describe the model to generate and evolve the pulsars, we describe the model to detect the pulsars in the simulation in Sec.~\ref{sec:detection}, the results are presented in Sec.~\ref{sec:results}, discussed in Sec.~\ref{sec:discussion} and summarized in Sec.~\ref{sec:conclusions}. 

\section{Evolution model} \label{sec:model}

Compared to previous studies, the novelty of the present work lies in the careful treatment of the evolution of binary systems up to the formation of compact objects in their final stage. At the beginning of the simulation, all of the secondary stars considered in the binaries are main sequence (MS) stars that are at a certain percentage of their lives. This percentage is taken from a random uniform distribution between 0 and 100 \%, while the NS in the binary starts at age 0. We begin by presenting the stellar evolution code, Stellar EVolution for N-body (SEVN), followed by a description of the binary processes it includes. We then introduce the adopted birth distributions and initial characteristics, and outline the prescriptions implemented for the evolution of neutron stars. Next, we discuss the spatial evolution of the systems, before finally addressing the concept of the death valley.
%and the detectability of our simulated pulsars in both radio and $\gamma$-ray wavelengths.  

\subsection{The SEVN code}\label{subsec:sevn}
SEVN is a state-of-the-art binary population synthesis code. It models single stellar evolution by interpolating pre-computed stellar tracks on the fly, and accounts for binary interactions using analytic and semi-analytic prescriptions \citep{sm17, smg+19, msm+20, sim+23}. In this work, we use the version of SEVN described in \citet{imc+23}\footnote{\url{https://gitlab.com/sevncodes/sevn}}. In subsections \ref{subsec:binary_processes}, \ref{subsec:birth_prop}, \ref{subsec:NS_evo} and \ref{subsec:spatial_evo}, we detail a general overview of the main features that we use from SEVN, with modifications on the NS evolution proposed by SEVN to get the best description possible. We refer to \citet{imc+23} for a detailed description of the code.  
SEVN require us to choose the maximum mass of a neutron star (NS) that can be reached in the simulation. Thus, we set this maximum to be 3~$M_{\odot}$ as in the work of \citet{sim+23}. 

\subsection{Binary processes} \label{subsec:binary_processes}
Various binary processes come into play during the course of binary evolution, in the following we describe the ones relevant in this work that are implemented in SEVN. The Roche lobe $R_L$, of a star in a binary system defines the region of space within which matter remains gravitationally bound to the star. When a star fills its Roche lobe, material can flow toward the companion star under the influence of its gravitational pull, a process known as Roche lobe overflow (RLO). RLO affects several key parameters, including the mass ratio, the individual masses and radii of the stars, and the binary’s semi-major axis. At each time step, SEVN evaluates the Roche-lobe radii of both stars using the analytical formula derived by \citet{e83}
\begin{equation} \label{eq:roche_lobe}
\frac{R_L}{a} = \frac{0.49 \ q^{2/3}}{0.6 \ q^{2/3} + \ln\left(1+q^{1/3}\right)},
\end{equation}
If the radius of the companion of the NS satisfy $r \geq R_L$, then a RLO episode starts and mass falls from the secondary star, which filled its Roche lobe, to the NS (the opposite case is also possible, although very unlikely as the system would need to be very compact). The mass transfer can then be either stable or unstable, in order to decide in which case the system is, in SEVN when the mass ratio $q$ is greater than a critical value $q_c$ that depends on the stellar evolutionary phase, the mass transfer is unstable on a dynamical time-scale. The values of $q_c$ used are the same as in \matteo{\citet{htp02,smg+19,gm20}}, see \matteo{the values in the first column of} table 3 in \citet{imc+23}. %According to this model, mass transfer will always be stable if the donor star is a MS star or in the Hertzprung gap evolutionary phase. 

A stable mass transfer results in the mass-loss rate, $\dot{M}_d$, proposed by \citet{htp02}
\begin{equation} \label{eq:mlossrate}
\dot{M_d} = -F(M_d) \left(\ln\frac{R_d}{R_{L,d}}\right)^3 \text{M$_{\odot}$yr$^{-1}$}
\end{equation}
where $F(M_d)$ is a normalization factor, $R_d$ is the radius of the donor and $R_{L,d}$ is the Roche lobe of the donor star. SEVN accounts for non-conservative mass transfer, meaning that the donor star can lose more mass than the companion actually accretes. The accreted mass, $\dot{M}_a$, is then modeled according to the following prescription
\begin{equation}
\dot{M_a}=
\begin{cases}
\text{min}\left(\dot{M}_{\rm Edd},-f_{\rm MT} \dot{M_d}\right) \text{ if the accretor is a compact object}\\
-f_{\rm MT} \dot{M_d} \ \ \ \ \ \ \ \ \ \ \ \ \ \ \ \ \ \ \ \ \ \ \ \ \text{otherwise,}
\end{cases}
\end{equation}
where $\dot{M}_{\rm Edd}$ is the Eddington accretion rate and $f_{\rm MT}$ is the mass accretion efficiency, which is within the range [0,1]. In SEVN, $f_{\rm MT} = 0.5$ \citep{bms+21,imc+23}.

In the case of unstable mass transfer, the binary either results in a merger or a common envelope (CE). There are debates in the community about if it is possible to have spin-up during the CE phase \citep{obg+11,mr15,cfb+18,csh+20}. Because of the controversy, spin-up during CE is not included in SEVN. Therefore, if the mass transfer is not stable, in this simulation the pulsar can not be recycled, that is why we do not describe the CE evolution here, but see \citet{imc+23,sim+23} for more details. 

\subsection{Birth properties} \label{subsec:birth_prop}
Many birth characteristics generated for each pulsar are independent of their companion, such as the position in the Galaxy ($x_0$,$y_0$,$z_0$), inclination angle ($\chi_0$), spin period ($P_0$), magnetic field ($B_0$). Therefore, these quantities are generated from the same distributions, with the same parameters, as described in the paper of \citet{spm+24}. However\matteo{, compared to our study of the normal pulsars,} we did change two elements in this work: firstly, the kick velocity distribution was modified by adopting a lower dispersion of $\sigma_v = 70$\,km\,s$^{-1}$ \matteo{which better matches the recycled pulsar population and is consistent with earlier works \citep{sgh07,ghf+18}. Indeed, this prescription is} to reflect the assumption that the binary survived the supernova explosion of the primary star that formed the neutron star. All simulated systems are thus assumed to remain bound after the first supernova event \matteo{that initially formed the neutron star in the binary}. %We do not consider the disrupted systems at this stage, as they would most likely evolve into isolated neutron stars.
\matteo{Furthermore, this prescription provides a better match to the recycled pulsar population, as a high velocity dispersion would result in the loss of a large fraction of systems (resulting in $\sim$~3.5 times fewer detectable neutron stars), thereby requiring a significantly higher birth rate to reproduce the observed population.} Nevertheless, if the secondary star undergoes a supernova explosion \matteo{to become a neutron star or a black hole} later in the evolution, the binary may still be disrupted due to the associated mass loss and natal kick \matteo{from the formed neutron star or black hole}, as implemented in the SEVN code. Secondly, instead of using the radial distribution in the galactic plane of \citet{yk04}, we used a truncated normal distribution (the distribution is truncated to be always greater than 0) with a mean $\mu_{R}=11.3$~kpc, and a standard deviation $\sigma_R = 1.8$~kpc \matteo{to have the neutron stars born in the spiral arms (see \citet{spm+24} to check how the neutron stars are then placed into the spiral arms)}. Actually using the radial distribution proposed by \citet{yk04} allowed to detect too many pulsars in the galactic center compared to observations. That is why we opted for the truncated normal distribution \matteo{for the radial distribution in the spiral arms,} which was also an option in \citet{fk06,tl25}. 

The population of NS generated have their ages taken from a random uniform distribution between a minimum age of $5\times10^7$~yr, as it is very unlikely that a pulsar can be recycled within a shorter time, and a maximum age of $13.9\times10^9$~yr, which is within the range of the estimated age of the universe \citep{rfc+98}. In each simulation, the number of sources is characterized by a birth spacing $X$, where $X$ is chosen as an integer number of years between the births of subsequent pulsars. In this paper, we report this quantity as either the birth spacing $X$ in years or the birth rate $R_{\rm recycled} =  (1/X)\,{\rm yr}^{-1}$. This approach can be used to verify whether a constant birth rate is realistic, by comparing the results from the simulation to observation. Before running the simulation, we choose the number of binaries, and then by computing $R_{\rm recycled} = \frac{\rm Recycled \ pulsars \ formed}{\rm age_{max} - age_{min}}$ at the end of the simulation, we get the birth rate. In this work, simulating between \matteo{$4.6\times10^5$ and $5.7\times10^5$} pulsars in a binary allowed us to get similar $P-\dot{P}$ diagram between observations and simulations, resulting in a birth rate between \matteo{$\rm 3.3-4.1 \times10^{-6} \ yr^{-1}$.} We discuss the birth rate range obtained in Sect.~\ref{sec:discussion}. 

The mass of the NS progenitor $M_1$ is drawn from Kroupa's initial mass function (IMF) \citep{k01}, within the range between \matteo{[8-25]}~$M_{\odot}$ \citep{bt08}
\begin{equation} \label{eq:distrib_mass_progenitorNS}
P(M_1) \propto M_1^{-2.3}.
\end{equation}
Although our simulation starts with the NS already formed, meaning we are ultimately interested in its initial mass, we still need to know the progenitor mass in order to determine the mass of the companion star. This is because the mass ratio of the binary $q = \frac{M_2}{M_1}$, with $M_2$ being the mass of the secondary star, is drawn from a power-law distribution as suggested by \citet{smk+12}
\begin{equation} \label{eq:distrib_q}
    P(q) = q^{-0.1}.
\end{equation}
Then, we compute $M_2$ by knowing $q$ and $M_1$. Each pulsar has its birth mass $M_{\rm NS,i}$ drawn from a gaussian distribution such as
\begin{equation} \label{eq:distrib_MNS_init}
P(M_{\rm NS,i}) = \frac{1}{\sigma_M\sqrt{2\pi}} e^{-(M_{\rm NS,i} - \mu_M)^2/(2\sigma_M^2)},
\end{equation}
where $\mu_M = 1.33 \ M_{\odot}$ and $\sigma_M = 0.09 \ M_{\odot}$. This distribution is derived from a fit of the Galactic binary neutron stars (BNS) masses \citep{opn+12,of16}. \matteo{Our independent sampling of progenitor mass and neutron-star mass removes the statistical correlation whereby more massive progenitors tend, on average, to produce more massive neutron stars \citep{sew+16}. This may influence the neutron star mass distribution obtained in this work.} MSPs are very fast rotating NSs, therefore for $P < 6.5$~ms, we can not neglect anymore the spin-down caused by gravitational waves, see Appendix \ref{appendix:AppA} for more details. We can define a triaxial inertia tensor which is diagonalized in the basis of its principal axes. These axes are denoted by $\boldsymbol{e}_1$, $\boldsymbol{e}_2$, $\boldsymbol{e}_3$ with the corresponding moments of inertia given by $I_1=I$, $I_2=I\left(1+\epsilon_{12}\right)$ and $I_3=I\left(1+\epsilon_{13}\right)$ respectively. $\epsilon_{12}$ and $\epsilon_{13}$ characterize the ellipticiy, i.e., the degree of deviation from perfect stellar sphericity. However, in our case we will only consider as if we had biaxial stars, which is sufficient for our purposes, as considering triaxial stars would significantly increase the complexity without providing additional physical insight beyond the key requirement that the stars is deformed, therefore we have $\epsilon_{12}=0$ and $\epsilon_{13}=\epsilon \neq 0$. Thus, for each NS, we draw its ellipticity $\epsilon$ from a log uniform distribution between [-13,-5], in order to have $\epsilon$ between [$10^{-13},10^{-5}$]. These limits are suggested by the study of \citet{gc22}. 

SEVN also requires the metallicity of the secondary star. As it would necessitate to take into account the environment and the past of the binaries to have an accurate value taken for the metallicity, no assumption is made about it. The metallicity $Z$ is taken from a random log-uniform distribution between [0.0002, 0.02]. Furthermore, still concerning the secondary star, SEVN also needs the initial ratio $\Omega_s/\Omega_c$, which is the spin frequency of the star over its critical spin frequency, we take it from a uniform distribution between [0,1]. We chose a uniform distribution over all possible values of this ratio, as we found that the choice of distribution had little influence on the results.

The binary eccentricity~$e$ distribution is derived from a power-law as suggested by the study of \citet{smk+12} where they found this distribution for massive binaries 
\begin{equation} \label{eq:distribution_e}
P(e) = e^{-0.42}.
\end{equation}
The semi-major axis $a$ of the binary is drawn from a distribution that is flat in $\log(a)$ for wide binaries, and decreases smoothly at close separations, as proposed by \citet{hgc+20}, 
\begin{equation} \label{distrib_semi_major_axis}
an(a) =
\begin{cases}
\psi_{\text{sep}} \left( \dfrac{a}{a_0} \right)^m & \text{if } a \leq a_0 \\
\psi_{\text{sep}} & \text{if } a_0 < a < a_1,
\end{cases}
\end{equation}
where $\psi_{\rm sep} \approx 0.070$, $a_0=10 R_{\odot}$, $a_1 = 5.75 \times 10^6 R_{\odot}$, $R_{\odot} \approx 696340$~km and $m\approx1.2$. \matteo{We do not claim that these distributions represent the exact initial distributions for NS+MS binaries, which are currently unknown. Nevertheless, they cover the relevant parameter space for $a$ and $e$, and using uniform distributions over the same ranges does not reproduce the observed population as well. Therefore, we retained the distributions of $a$ and $e$ for massive binaries, even though they may not accurately reflect reality.}
The angle between the rotation axis of the NS and the normal to the orbital plane, which we call $\alpha$, the spin-orbit angle is assumed to follow an isotropic distribution generated from a uniform distribution $U\in[0,1]$ and given by $\alpha = \arccos (2\,U-1)$. 

\subsection{Neutron star evolution} \label{subsec:NS_evo}
\subsubsection{General evolution model}\label{subsubsection:gen_evo_model}
Because SEVN evolves pulsars in vacuum as they spin down, we changed the evolution proposed by SEVN with a force-free model evolution taking into account the plasma in the magnetosphere. Further, we adapted the model from \citet{ba21} to also take into account the gravitational waves (GWs) emission if the NS reaches a spin frequency $\Omega \geq 960$~rad.s$^{-1}$. As soon as the NS reaches again $\Omega < 960$~rad.s$^{-1}$, the GWs are neglected. While a fully accurate treatment would require modeling the impact of gravitational waves on the inclination and spin-orbit angle evolution, we neglect these effects as they would greatly complicate the model, while most of the time the GWs emission can still be neglected. Thus, in this work the evolution of a NS can be described by this set of differential equations
\begin{subequations}
\begin{align}
\label{eq:diff_eq_model1}
&I \, \dot{\Omega} = n_1\cos\alpha + n_2 + n_3 \left(1+\sin^2\chi\right) + n_{\rm GW} - \dot{I}\Omega, \\
\label{eq:diff_eq_model2}
&I \,  \Omega \, \dot{\alpha} = -n_1 \sin\alpha, \\
\label{eq:diff_eq_model3}
&I \, \Omega \, \dot{\chi} = \eta A\left(\eta,\alpha,\chi\right) n_1 \sin^2\alpha \cos\alpha \sin\chi \cos\chi + n_3\sin\chi \cos\chi,
\end{align}
\end{subequations}
where $I=\frac{2}{5}MR^2$ is the moment of inertia of the NS (we assume the spherical formula for I, although we consider an ellipticity $\epsilon$ for the NS, as this $\epsilon$ is very small, between $10^{-13}$ and $10^{-5}$). The torques $n_1, n_2, n_3, n_{\rm GW}$ are respectively, the averaged torques acting on the NS caused by accretion, magnetic braking due to magnetic field-disk interaction, pulsar's radiation loss and gravitational braking due to a non-null ellipticity. The coefficient $\eta$ is set to unity as suggested by \citet{ba21}, it describes the accretion torque modulation within the spin period. $A\left(\eta,\alpha,\chi\right)$ is a normalization function
\begin{equation} \label{eq:A_normalization}
A\left(\eta,\alpha,\chi\right) = \left[1- \frac{\eta}{2}\left(\sin^2\chi\sin^2\alpha+2\cos^2\chi\cos^2\alpha\right)\right]^{-1}
\end{equation}
The torques are defined by
\begin{subequations}
\begin{align}
\label{eq:accretion_torque}
&n_1 = \dot{M}_{\rm NS}\left(GM_{\rm NS} \ r_{\rm in}\right)^{1/2}, \text{ if } r_{\rm in} < r_{\rm co}, \\
\label{eq:B_disk_interaction_torque}
&n_2=-\frac{\mu^2}{3 r_{\rm co}^3}, \text{ if } r_{\rm in} < r_{\rm lc}, \\
\label{eq:n3_torque_pulsaradiationloss}
&n_3= -\frac{\mu^2}{r^3_{\rm lc}}, \\
\label{eq:n_gw}
&n_{\rm GW} = -\frac{4G I^2 \Omega^5 \epsilon^2}{5\pi^2c^5}, \text{ if } \Omega \geq 960 \text{ rad.s$^{-1}$}.
\end{align}
\end{subequations}
Here $G$ is the gravitational constant, $M_{\rm NS}$ is the mass of the NS, $\dot{M}_{\rm NS}$ is the mass accretion rate of the NS, $\mu^2=\frac{4\pi B^2 R^6}{\mu_0}$ is the magnetic moment with the permeability constant $\mu_0=4\pi \times 10^{-7}$~H/m, the radius of the NS, $R=12$~km, and the speed of light in vacuum $c=3\times 10^8$~m/s. The different radii defined $r_{\rm in}, r_{\rm co}$ and $r_{\rm lc}$ are respectively, the inner disk radius, corotation radius and light cylinder radius
\begin{subequations}
\begin{align}
\label{eq:r_in}
&r_{\rm in} = \xi \left(\frac{\mu^4}{2 G M_{\rm NS} \dot{M}_{\rm NS}^2}\right)^{1/7}, \\
\label{eq:r_co}
&r_{\rm co} = \left(\frac{GM_{\rm NS}}{\Omega^2}\right)^{1/3}, \\
\label{eq:r_lc}
&r_{\rm lc} = \frac{c}{\Omega},
\end{align}
\end{subequations}
where $\xi \sim 0.5$ is a dimensionless factor accounting for the detailed disk structure \citep{bzf+08}.

Since it is not straightforward to visualize the temporal evolution of a pulsar in the $P-\dot{P}$ diagram from this system of differential equations, we provide in the Subsection~\ref{subsec:evol_in_the_PPdot} an example of the typical evolutionary track of a pulsar that becomes a recycled pulsar.

\subsubsection{Spin-down} \label{subsubsec:spindown}
When the neutron star is not accreting matter, $n_1$ and $n_2$ are set to zero, and the star spins down as a result. Additionally, $n_{\rm GW}$ is also set to zero when $\Omega < 960$~rad.s$^{-1}$. When these conditions are satisfied, we can apply the simplified set of equations from \citet{spm+24}, which consider only the pulsar’s radiative losses. This approach allows for faster computation compared to solving the full system of equations \eqref{eq:diff_eq_model1}, \eqref{eq:diff_eq_model2}, and \eqref{eq:diff_eq_model3}. The equations are given by:
\begin{subequations}
\begin{align}
\label{eq:ffe_eq1}
\Omega \frac{\cos^2\chi}{\sin\chi}= \Omega_0 \frac{\cos^2\chi_0}{\sin\chi_0}, \\
\label{eq:ffe_eq2}
\ln(\sin \chi_0) + \frac{1}{2\sin^2\chi_0} + &K \Omega_0^2 \frac{\cos^4\chi_0}{\sin^2\chi_0} \frac{\alpha_d \tau_d B_0^2}{\alpha_d - 2} \left[\left(1+\frac{t}{\tau_d}\right)^{1-2/\alpha_d}-1\right] \notag \\ 
\phantom{{}={}}= \ln(\sin \chi) + \frac{1}{2\sin^2\chi},
\end{align}
\end{subequations}
where the quantities with a subscript 0 indicate their value at birth. Those without a subscript display their current value at present time. The time $t$ represents the true age of the pulsar and $K = 4\pi R^6/I \mu_0 c^3$. Equation \eqref{eq:ffe_eq1} is the integral of motion between $\Omega$ and $\chi$, the inclination angle, and equation \eqref{eq:ffe_eq2} allows to find the evolution of $\chi$ in a dipolar configuration with a magnetic field decay. This system was found by \citet{ptl14} for a spherically symmetric NS with a constant magnetic field and \citet{dpj22} adapted \eqref{eq:ffe_eq2} for a decaying prescription. Finally, when there is no accretion and $\Omega\geq 960$~rad.s$^{-1}$, we solve still \eqref{eq:diff_eq_model1} and \eqref{eq:diff_eq_model3}, with $n_1=n_2=0$.

\subsubsection{Spin-up} \label{subsubsec:spinup}
As mentioned earlier, the NS can spin-up only if it is accreting matter as $n_1$ is the only torque which is positive, therefore only $n_1$ can increase $\Omega$. Two conditions are necessary for the NS to accrete matter: first, the companion must reach its Roche lobe, then the NS needs to have its Keplerian angular velocity at the inner disk radius $\Omega_{{\rm K}|{r_{\rm in}}}$ greater than its co-rotation angular velocity $\Omega_{\rm co}$
\begin{equation} \label{eq:vdiff}
\Omega_{\rm diff} = \Omega_{{\rm K}|{r_{\rm in}}} - \Omega_{\rm co}.
\end{equation}
Thus, $\Omega_{\rm diff}>0$ is necessary for the NS to be accreting matter, the inner disk radius $r_{\rm in}$ is defined by equation \eqref{eq:r_in}.
When matter reaches the inner disk radius, the magnetic pressure dominates, forcing the material to follow the magnetic field lines and funnel onto the neutron star’s polar caps. However, if $\Omega_{\rm diff}<0$, the neutron star enters the propeller regime: the magnetic field lines at the magnetic radius rotate faster than the local Keplerian velocity, preventing accretion and expelling the matter away from the star \citep{abp24}. 

\subsubsection{Magnetic field evolution} \label{subsubsec:b_field_evol}
The evolution of neutron star magnetic fields remains a subject of debate. Some early pulsar population synthesis studies suggested that the magnetic field does not evolve over time \citep{bwh+92,fk06}. However, several physical mechanisms support the idea of magnetic field decay, including ohmic dissipation in the crust, ambipolar diffusion in the core, crustal vortex motion, internal turbulence, and energy loss through radiation \citep{sbm+90,caz04,dgp12,vrp+13}. A recent study that includes ohmic heating \citep{ip24}, along with period derivative measurements and spectral data from the neutron star RX J0720.4-3125, further suggests a rapid decay of its magnetic field. Moreover, more recent population synthesis studies have shown that models incorporating magnetic field decay provide a great match to the observed pulsar population \citep{gmv+14,spm+24}.
Therefore, we adopt a magnetic field decay following a power-law prescription 
\begin{equation} \label{eq:Bfield}
B(t) = B_0\left(1+\frac{t}{\tau_d}\right)^{-1/\alpha_d} +B_{\rm min},
\end{equation}
where $\alpha_d$ is a constant parameter controlling the speed of the magnetic field decay, $B_0$ the initial magnetic field, $B_{\rm min}$ the minimum magnetic field reachable by the neutron star, $t$ corresponds to the elapsed time since the birth of the neutron star and $\tau_d$ the typical decay timescale. We use $\alpha_d=1.5$ (as in \citet{spm+24} for the normal pulsars). Concerning $\tau_d$ we consider three different values, randomly chosen for a pulsar generated: 0.05, 1.4 and 2.3~Myr, with a probability of 0.77, 0.115 and 0.115 to have these values respectively. These three probabilities effectively approximate the probability density function of $\tau_d$ and were sufficient to reproduce the recycled pulsar population. Determining the full distribution of $\tau_d$ would require further investigation beyond the scope of this study. Physically, the $\tau_d$ distributions represent the range of possible evolutionary trajectories in the $P$–$\dot{P}$ diagram for recycled pulsars, similar to Fig. 10 in \citet{vrp+13}. Thus, each pulsar in the simulation follows one of three possible evolutionary paths, an empirical approach that better reproduces the $P$–$\dot{P}$ distribution than using a single path. These discrete values were selected purely by trial and error. Whether the NS is accreting or not, the magnetic field is decaying in this model, however accretion-induced field decay is also taken into account.

As the NS accretes matter from its companion, this matter goes along the magnetic field lines, to the polar caps. Progressively, the matter accreted can apply enough pressure to compress and bury the magnetic field under the NS surface \citep{pm07,wzc12}. This accretion-induced field decay process can be described by 
\begin{equation} \label{eq:acc_induced_field_decay}
B_{\rm AI}(t) = (B(t) - B_{\rm min}) \ e^{-\frac{\Delta M_{\rm NS}}{\Delta M_d}} + B_{\rm min},
\end{equation}
where $B(t)$ is computed via equation \eqref{eq:Bfield} to model the decay caused by the other physical mechanisms mentioned earlier, $\Delta M_{\rm NS}$ is the amount of accreted mass by the NS and $\Delta M_d$ is the magnetic field decay mass scale. Typically the accreted mass by a NS is less than 0.5~$M_{\odot}$ \citep{of16,yzl+25} and we adopt $\Delta M_d=0.025 \ M_{\odot}$ as in the work of \citet{csh+20} which is the optimal value found in their study.

To prevent the magnetic field of neutron stars from dropping to unrealistically low values, we impose for each pulsar a minimum magnetic field value $B_{\rm min}$, randomly drawn from a uniform distribution in the range [$10^3$, $10^4$]~T. This range is motivated by the study of \citet{zk06}, who investigated the lower limits that neutron star magnetic fields can realistically attain.

\subsection{Spatial evolution} \label{subsec:spatial_evo}
The motion of each binary's center of mass within the Galactic potential is evolved in the gravitational potential $\boldsymbol{\Phi}$. They are subject to an acceleration $\boldsymbol{\ddot{x}}$ given by
\begin{equation} \label{eq:acceleration_pulsar}
\boldsymbol{\ddot{x}} = - \boldsymbol{\nabla\Phi}.
\end{equation}
We integrate numerically equation \eqref{eq:acceleration_pulsar} with a Position Extended Forest Ruth-Like (PEFRL) algorithm \citep{omf02}, a fourth order integration scheme. Concerning the galactic potential, the Galaxy is divided in four distinct regions, which results in four different potentials: the bulge $\Phi_{b}$ and the disk $\Phi_{d}$ which have the form proposed by \citet{mn75}, the dark matter halo $\Phi_{h}$ which has the form proposed by \citet{nfw97} and the nucleus $\Phi_{n}$ represented by a Keplerian potential. The total potential of the Milky Way $\Phi_{tot}$ is the sum of these potentials:
\begin{equation} \label{tot_pot}
\Phi_{tot} = \Phi_{b} + \Phi_{d} + \Phi_{h} + \Phi_{n}.
\end{equation}
See \citet{spm+24} for a detailed analysis of the PEFRL scheme, its accuracy, and for the detailed expression of the galactic potentials used. 

Although SEVN does not integrate the full orbital motion of binaries with respect to their center of mass, it computes the evolution of the semi-major axis and eccentricity using prescriptions for mass exchange within the system, following equations (15) and (16) of \citet{htp02}. The dominant processes affecting these orbital parameters are tides, which are modeled through equations (34) to (36) of \citet{imc+23} (see the sections 2.3.1 and 2.3.4 of their paper for more details about wind mass transfer and tides respectively). In addition, SEVN accounts for the evolution of the semi-major axis and eccentricity driven by gravitational-wave emission. \citet{p64} describes in his work the orbital decay and circularization caused by GWs with the equation below
\begin{subequations}
\begin{align}
&\frac{da}{dt} = -\frac{64 \ G^3 m_1 m_2 (m_1+m_2)}{5 \ c^5 a^3 \left(1-e^2\right)^{7/2}} \left(1+\frac{73}{24}e^2 +\frac{37}{96}e^4\right), \\
&\frac{de}{dt} = -\frac{304 \ G^3 m_1 m_2 (m_1+m_2)}{15 \ c^5 a^3 (1-e^2)^{5/2}} \left(1+ \frac{121}{304}e^2\right),
\end{align}
\end{subequations}
where $G$ is the gravitational constant, $m_1$ and $m_2$ are the masses of the two stars in the binary, $c$ is the speed of light, $a$ is the semi-major axis and $e$ is the eccentricity of the orbit. In SEVN, these equations are solved when the GW merger time-scale, $t_{\rm merge}$ becomes shorter than the Hubble time. See appendix D of \citet{imc+23} for more details about the computation of the GW merger time-scale. 

\subsection{Death valley} \label{subsec:death_valley}
\begin{figure}[h]
\resizebox{\hsize}{!}{\includegraphics{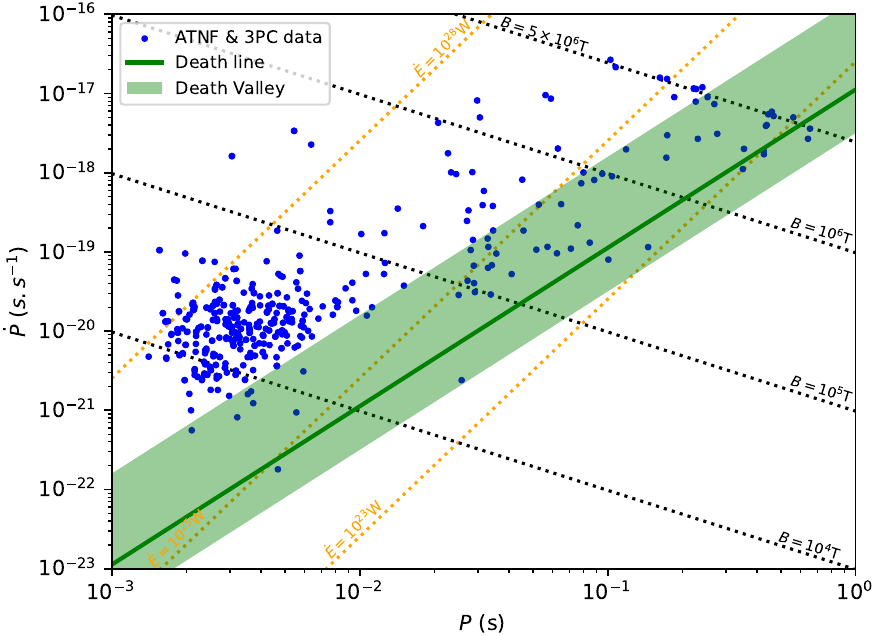}}
\caption{$P-\dot{P}$ diagram of the observed pulsars considered in this work along with the death line, green solid line, and death valley, shaded green area.}
\label{fig:death_valley}
\end{figure}

Spread in the parameter values of the model causes significant variations in the death line, and thus a death valley rather than a single death line describes the condition for pulsar extinction in the $P-\dot{P}$ diagram. Nonetheless, this death valley must be based on a death line definition, we choose to extend the death line from \citet{mbm+20}, originally defined for normal pulsars to access the criteria for generation of coherent radio emission given as follows.
\begin{equation}
\label{eq:death_line}
\dot{P}_{\rm line} = \frac{3.16 \times 10^{-19} \ T_6^4 \ P^2}{\eta^2 b \cos^2 \chi_l} \ .
\end{equation}
Here, $T_6 = T/10^6$~K, where $T$ corresponds to the surface temperature of the polar cap and $P$ is the spin period of the pulsar in s. The parameter $\eta = 1 - \rho_i/\rho_{GJ}$ is the electric potential screening factor due to the ion flow, where $\rho_i$ corresponds to the ion charge density and $\rho_{GJ}$ to the Goldreich-Julian charge density above the polar cap. The quantity $b$ is the ratio of the actual surface magnetic field to the dipolar surface magnetic field and $\chi_l$ is the angle between the local magnetic field and the rotation axis. 
% Pas de phrase isolé
In order to obtain the death valley in Fig.\ref{fig:death_valley}, the parameters described above are drawn from distributions, found and described in \citet{spm+24}. 

\subsection{Evolution in the $P-\dot{P}$ diagram of a recycled pulsar} \label{subsec:evol_in_the_PPdot}

\begin{figure}[h]
\resizebox{\hsize}{!}{\includegraphics{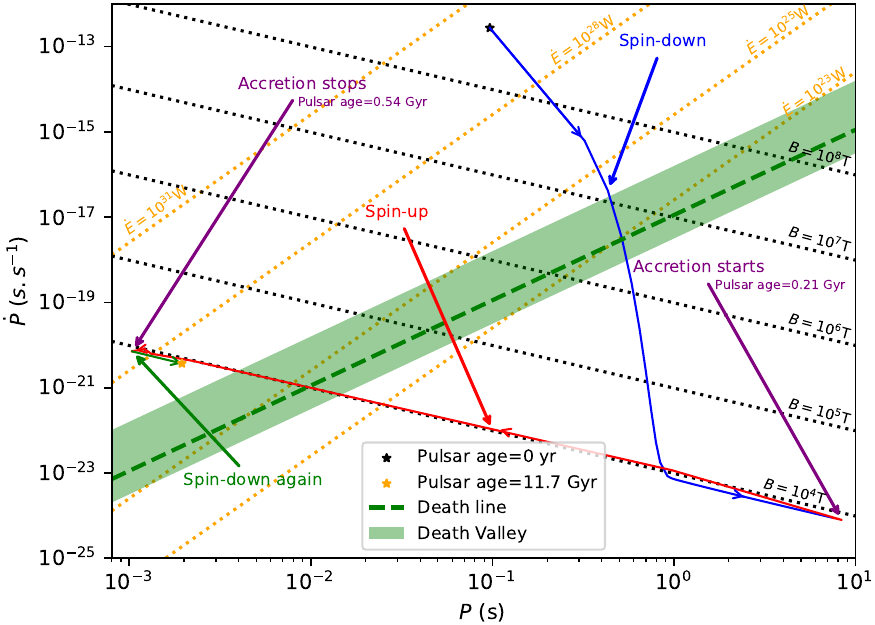}}
\caption{Evolution in the $P-\dot{P}$ diagram of a pulsar in the simulation which becomes a MSP.}
\label{fig:PPdot_evol_one_pulsar}
\end{figure}

In order to better visualize how a pulsar \matteo{evolve in the $P-\dot{P}$~diagram to become a} recycled pulsar, Fig.~\ref{fig:PPdot_evol_one_pulsar} shows the track followed by a pulsar evolving in the simulation, starting with an initial period \matteo{$P_0=9.7\times10^{-2}$~s, period derivative $\dot{P}_0=2.7\times10^{-13}$~s/s, and mass $M_{\rm NS,0} = 1.38\ M_{\odot}$, its companion being a $\sim8 \ M_{\odot}$ main sequence star. Since the mass ratio of these two stars, $q=\frac{M_2}{M_{\rm NS}}=5.8$, exceeds the critical mass ratio for main sequence stars in the simulation, $q_c = 3.0$, it means that the mass transfer would be unstable. This implies that a CE phase must have occurred, allowing the secondary star to lose enough mass in the envelope, which was then ejected, for the system to later enter a stable mass-transfer phase. For further details on binary evolution, see \citet{th23}.} We can basically distinguish three major phases in the evolution of this pulsar: first, the pulsar was dominated by the spin-down (caused mostly by the torque $n_3$ presented in \ref{subsec:NS_evo}) and evolved towards a spin period close to 10~s. Then, when the accretion started \matteo{(after 0.21~Gyr of evolution)} we can clearly see that the pulsar, which passed the death line before, comes back from the commonly called graveyard of pulsars to have a very rapid rotation, close to the millisecond, thanks to the spin-up (caused by the torque $n_1$ presented in \ref{subsec:NS_evo}) that dominates this phase. Finally, when the accretion ends \matteo{(the accretion lasted $\sim 0.33$~Gyr)}, the pulsar can continue to evolve by being dominated once again by spin-down. However, compared to when the pulsar was young, the spin-down is much weaker, due to the reduced value of the magnetic field, which is now much lower than at its birth\matteo{, that is why $\sim10$~Gyr after the end of the accretion phase, the pulsar barely moved in the $P-\dot{P}$ diagram}. \matteo{The pulsar finished its evolution with $P=1.9\times10^{-3}$~s, $\dot{P}=1.1\times10^{-20}$~s/s, $M_{\rm NS} = 2.3\  M_{\odot}$ (thus, the pulsar accreted $\sim0.9 \ M_{\odot}$) and ended up isolated as its companion was ejected after its supernova explosion. }
%its companion became a COWD of $\sim0.7 \ M_{\odot}$.} 

\section{Detection model} \label{sec:detection}
In this section, we address the detectability of our simulated pulsars in both radio and $\gamma$-ray wavelengths. First of all, in this work we considered exclusively two radio surveys: the FAST Galactic Plane Pulsar Snapshot survey (FAST GPPS, \citet{hww+21}) or the Parkes Multibeam Pulsar Survey (PMPS, \citet{mlc+01}) as they detected the most MSPs. While for the $\gamma$-ray wavelengths we are interested in the comparison with the detections obtained by the Fermi/LAT instrument, therefore a comparison with the Third Fermi Large Area Telescope Catalog (3PC). 

For each pulsar, we verified whether it met the detection criteria based on three factors. The first is the beaming fraction, which represents the portion of the sky swept by the radiation beam and depends on the wavelength (radio or $\gamma$-ray), the spin rate, geometry, and emission region. Before detailing how the radio and $\gamma$-ray beaming fractions are computed, we first define the relevant angles.

The angle between the line of sight and the rotation axis is $\zeta = (\widehat{\vec{n_{\Omega}},\vec{n}})$, where $\vec{n_\Omega} = {\vec{\Omega}}/{||\vec{\Omega}||}$ is the unit vector along the rotation axis, and $\vec{n}$ is the unit vector along the line of sight. The inclination angle between the rotation axis and the magnetic moment is $\chi = (\widehat{\vec{n_{\Omega}},\vec{\mu}})$, with $\vec{\mu}$ the unit vector along the magnetic moment. Finally, the impact angle $\beta = (\widehat{\vec{\mu},\vec{n}})$ is related by $\chi + \beta = \zeta$.

We chose an isotropic distribution for the Earth viewing angle, $\zeta$, as well as for the orientation of the unitary rotation vector. The Cartesian coordinates of the unit rotation vector, $\vec{n_{\Omega}}$, are $(\sin\theta_{n_{\Omega}} \cos\phi_{n_{\Omega}}, \sin\theta_{n_{\Omega}} \sin\phi_{n_{\Omega}}, \cos\theta_{n_{\Omega}})$. We set the Sun's position at $(x_{\odot}, y_{\odot}, z_{\odot}) =$ (0~kpc, 8.5~kpc, 15~pc) \citep{s19}. The coordinates for $\vec{n}$ are
\begin{equation} \label{eq:coord_n_vector}
\vec{n} = \left(\frac{x-x_{\odot}}{d},\frac{y-y_{\odot}}{d},\frac{z-z_{\odot}}{d} \right) .
\end{equation}
To compute the pulsar distance from Earth, we use the formula for the distance,
\begin{equation} \label{eq:dist}
d = \sqrt{(x-x_{\odot})^2 + (y-y_{\odot})^2 + (z-z_{\odot})^2 } .
\end{equation}
\subsection{Radio detection model} \label{subsubsec:radio_detection}
The beaming fraction in radio depends on the half opening angle of the radio emission cone,~$\rho$. Since our selected sample of observed pulsars includes a few pulsars with $P>0.1$~s, the half opening angle $\rho$ of simulated pulsars that reach a similar value is computed according to \citet{lk04} by
\begin{equation} \label{eq:rhoangle}
\rho = 3 \sqrt{\frac{\pi \, h_{\rm em}}{2 \, P \, c}}
,\end{equation}
where $h_{\rm em}$ is the emission height, $P$ the pulsar spin period, and $c$ the speed of light. The emission height is assumed constant with an average value of $h_{\rm em} = 3\times10^5$ m, based on observational estimates from a sample of pulsars \citep{wj08, m17, jk19, jkk+23}. The cone half-opening angle $\rho$, as defined in Eq.~\eqref{eq:rhoangle}, applies only to the last open field lines of a dipolar magnetic field and is valid primarily for slow pulsars, where the emission altitude is high enough for higher-order multipolar components to be negligible. For fast pulsars, however, relativistic effects due to rapid rotation become significant and must be taken into account \citep{xkj+98}. Furthermore, little is known about the characteristics of the half-opening angle and the emission height in fast pulsars. For this reason, and since we adopt a purely dipolar geometry for all pulsars, even though multipolar components are expected to play a role in reality, we retain a formula similar to Eq.~\ref{eq:rhoangle}, following the prescription of \citet{kg98,kg03} for $\rho$:
\begin{subequations}
\begin{align}
\label{eq:rho_MSP}
\rho &= 1.24^{\circ} \, r_{\rm alt}^{1/2} \, P^{-1/2}, \\
\label{eq:rkg}
f(r_{\rm alt}) &= \frac{1}{\sigma_{\rm alt}\sqrt{2\pi}}  e^{-\left(r_{\rm alt} - \mu_{\rm alt} \right)^2 / \left(2\sigma^2_{\rm alt}\right)},
\end{align}
\end{subequations}
where $r_{\rm alt}$ denotes the emission altitude in units of neutron star radii, and $\mu_{\rm alt}$ and $\sigma_{\rm alt}$ are respectively the mean and standard deviation of the Gaussian distribution ($f(r_{\rm alt})$) from which $r_{\rm alt}$ is drawn.
Running multiple simulations showed that adopting $\mu_{\rm alt} = 8\,R_{\rm NS}$ and $\sigma_{\rm alt} = 4\,R_{\rm NS}$ provides the best agreement with observations, meaning that both radio pulse profiles and the fraction of $\gamma$-ray pulsars that could be visible in radio are consistent with the observations. 
%In this case, we adopt the prescription from \citet{kg98, kg03}, given by
%\begin{subequations}
%\begin{align}
%\label{eq:rho_MSP}
%\rho &= 1.24^{\circ} \, r_{\rm KG}^{1/2} \, P^{-1/2}, \\
%\label{eq:rkg}
%r_{\rm KG} &= 120 \, \nu^{-0.26}_{\rm GHz} \, \dot{P}_{-15}^{0.07} \, P^{0.3},
%\end{align}
%\end{subequations}
%where $r_{\rm KG}$ represents the altitude of the emission in stellar radii, also $\dot{P}_{-15}=\dot{P}/10^{-15}$, $\nu_{\rm GHz}$ is the survey frequency in gigahertz. \matteo{We increased the emission altitude compared to \citet{kg98,kg03} by a factor of three, as this provides a more satisfactory agreement with the fraction of $\gamma$-ray pulsars that would also exhibit radio emission between observations and simulations. Although, using a value of $r_{\rm KG}$ multiplied by some factor still leads to a consistent population, constraining the exact height of the radio emission through population synthesis seems challenging.}

A pulsar can be detected in radio if $\beta = |\zeta - \chi| \leq \rho$ or $\beta = |\zeta - (\pi - \chi)| \leq \rho$, corresponding to visibility from the northern or southern magnetic hemisphere, respectively. 
%In addition, the inclination angle $\chi$ must satisfy $\chi \geq \rho$ and $\chi \leq \pi - \rho$ to ensure that the line of sight intersects the emission cone, an essential condition to observe pulsations.
An additional useful quantity is the radio pulse width, $w_r$, which is computed following \citet{lk04} as
\begin{equation} \label{eq:observedwidthprofile}
\cos(\rho) = \cos(\chi)\cos(\zeta) + \sin(\chi)\sin(\zeta)\cos(w_r/2) .
\end{equation}

For radio flux density, we use a formula similar to that in \citet{jsk+20}, designed for 1.4 GHz observations by the Parkes telescope in the southern Galactic plane \citep{kbm+03,lsf+06,ccb+20} and Arecibo in the northern plane \citep{cfl+06}. The main difference in this work lies in the power-law dependence on the spin-down luminosity $\dot{E}=-I\Omega\dot{\Omega}$. \citet{jk17} found that using $\dot{E}^{0.5}$ overpredicted young pulsars, while $\dot{E}^0$ overrepresented old ones; they adopted $\dot{E}^{0.25}$ as a compromise. In our case, we found that lowering the exponent to $0.125$ improved agreement with observations. Thus, the flux density formula is
\begin{equation} \label{eq:rad_lum}
F_{\rm\! r} = 9 \ \text{mJy} \left(\frac{d}{1 \ \text{kpc}}\right)^{-2} \left(\frac{\dot{E}}{10^{29} \ \text{W}}\right)^{0.125} \times 10^{F_{\rm j}}
,\end{equation}
where $d$ is the distance in kpc and $F_{\rm j}$ is the scatter term which is modeled as a Gaussian with a mean of 0.0 and a variance of $\sigma$ = 0.2. The detection threshold in radio is set by the signal to noise ratio defined by
\begin{equation} \label{eq:SNratio}
S/N = \frac{F_{\rm\! r}}{S_{\rm survey}^{\rm min}} .
\end{equation}  
We directly compute the radio flux, without computing the luminosity in radio that overlooks the fact that the luminosity received will depend on the geometry of the beam. A pulsar is detected in radio if its signal-to-noise ratio (S/N) exceeds the threshold imposed by FAST GPPS \citet{hww+21} or PMPS \citet{mlc+01}. Indeed, in this work we focus exclusively on these two radio surveys. The region of observations of these surveys are taken into account, see Table \ref{Table:survey_params}. S/N depends on $S_{\rm survey}^{\rm min}$ which is the minimum detectable flux, which is related to the instrumental sensitivity, the period of rotation $P$ and the observed pulse profile width of radio emission $\Tilde{w}_r$. To compute the observed pulse profile width, we adopt the same formula as in \citet{cm03}.
\begin{subequations} 
\begin{align}
&\Tilde{w}_r = \sqrt{\left( {w_r\,P}/{2\pi}\right)^2 + \tau_{\rm samp}^2 + \tau_{\rm DM}^2 + \tau_{\rm scat}^2} \label{eq:pulse_profile_ISM_scatt}, \\
&S_{\text{survey}}^{\text{min}} = \frac{C_{\rm thres} [T_{\rm sys}+ T_{\rm sky}(l,b)]}{G \sqrt{N_p \ B \ t }} \sqrt{\frac{\Tilde{w}_r}{P-\Tilde{w}_r}} \label{eq:sensitivity}.
\end{align}
\end{subequations}

Equation~\eqref{eq:pulse_profile_ISM_scatt} accounts for interstellar medium (ISM) density fluctuations, including dispersion ($\tau_{\rm DM}$) and scattering ($\tau_{\rm scat}$) caused by interactions with free electrons in the Galaxy during the radio pulse's propagation. Instrumental effects are also considered through the sampling time, $\tau_{\rm samp}$. More details on these parameters are provided in Appendix \ref{appendix:AppB}.

In Equation~\eqref{eq:sensitivity}, where we detail the expression of the minimum detectable flux, $C_{\rm thres}$ is the detection threshold of the survey, $T_{\rm sys}$ is the system temperature in K, $G$ is the telescope gain in K.Jy$^{-1}$, $N_p$ is the number of polarizations, $B$ is the total bandwidth in Hz and $t$ is the integration time in s. The quantities which have just been presented, are the survey characteristics, which depend on whether we consider FAST GPPS or PMPS (see Table \ref{Table:survey_params}). $T_{\rm sky}$ is the sky background temperature in K, at the longitude $l$ and latitude $b$. \citet{rdb+15} provided a refined version of the temperature map of \citet{hss+82}. Since the data were obtained at 408 MHz, in order to rescale to the correct observing frequencies, a power-law dependence is assumed with the form $T_{\rm sky} \propto f^{-2.6}$ \citep{jlm+92}. We use Equation~\eqref{eq:sensitivity} to compute the minimum flux required for radio detection. 
\begin{table}[h]
\caption{Survey parameters of the Parks Multibeam Pulsar Survey (PMPS) and the FAST Galactic Plane Pulsar Snapshot survey (FAST GPPS).} 
\label{Table:survey_params} 
\centering 
\begin{tabular}{c c c} 
\hline\hline 
Survey & PMPS & Fast GPPS \\
\hline
Sky region & -100°< $l$ < 50° & -180° < $l$ < 180°\\
& |$b$| < 5° & |$b$| <10°\\
$f$ (GHz) & 1.374 & 1.0 - 1.5\\
$\Delta f_{ch}$ (kHz)& 3000 & 244 \\
$\tau_{\rm samp}$ (µs) & 250 & 50 \\
$G$ (K.Jy$^{-1}$) & 0.735 & 16\\
$N_p$ & 2 & 2 \\
$B$ (MHz) & 288 & 450\\
$t$ (s) & 2100 & 300\\
$T_{\rm sys}$ & 21 & 25\\
$C_{\rm thres}$ & 9 & 9\\
\hline 
\end{tabular}
\end{table}

%In addition we only use the parameters of PMPS to compute $\tau_{DM}$, we could have also used the parameters of The Pulsar Arecibo L-band Feed Array (PALFA) Survey; however, the differences in the results in the end are not large if we use these parameters. Consequently, it continues to serve as a favorable approximation to exclusively utilise the parameters delineated by PMPS.

\subsection{$\gamma$-ray detection model} \label{subsubsec:gamma_detection}
The $\gamma$-ray emission model is based on the striped wind scenario, where $\gamma$-ray photons are produced in the current sheet of the pulsar wind. Detection requires the observer’s line of sight to lie near the equatorial plane, with the inclination angle satisfying $|\zeta - \pi/2| \leq \chi$.
The $\gamma$-ray luminosity is taken from the study by \citet{khk+19}, who found it follows a fundamental plane relation. Their full 3D model depends on the magnetic field $B$, the spin-down luminosity $\dot{E}$, and the cut-off energy $\epsilon_{\rm cut}$. Since our PPS does not include $\epsilon_{\rm cut}$, we adopt the simplified 2D version of their model,
\begin{equation} \label{eq:gamma_lum}
L_{\gamma(2D)} = 10^{26.15 \pm 2.6} \ W \ \left(\frac{B}{10^8 \ \text{T}}\right)^{0.11\pm 0.05} \ \left(\frac{\dot{E}}{10^{26} \ \text{W}}\right)^{0.51 \pm 0.09},
\end{equation}
$\dot{E}$ is the spin-down luminosity and the associated $\gamma$-ray flux detected on earth is computed with 
\begin{equation} \label{eq:flux_gamma}
F_{\gamma} = \frac{L_{\gamma(2D)}}{4\,\pi \, f_{\Omega}\, d^2}
,\end{equation}
where $f_{\Omega}$ is a factor depending on the emission model reflecting the anisotropy. For the striped wind model, \citet{p11} showed that this factor varies between 0.22 and 1.90. In this work, we use an updated version of Fig.~7 of \citet{p11} which allow to obtain $f_{\Omega}$ knowing the viewing angle $\zeta$ and $\chi$ the inclination angle. Therefore, this update allows to get more realistic values for $f_{\Omega}$ between 0.13 and 316. See Appendix \ref{appendix:AppC} for more details, using the force-free model. 
%Nevertheless, an approximation is made : if $\chi < - \zeta + 0.6109,$ then $f_{\Omega}=1.9;$ otherwise $f_{\Omega} = 1$. As can be observed in Fig. 7 of \citet{p11}, $f_{\Omega}$ is usually equal to one of this two values, depending on the value of the inclination angle, $\chi$; hence this approximation.
The pulsar is detected in gamma depending on the instrumental sensitivity as described below. 

The $\gamma$-ray sensitivity is determined based on the expected performance of the Fermi/LAT instrument\footnote{\url{https://fermi.gsfc.nasa.gov/ssc/data/analysis/documentation/Cicerone/Cicerone_LAT_IRFs/LAT_sensitivity.html}}. For each simulated pulsar, we consult the all-sky sensitivity map provided by Fermi/LAT\footnote{\url{https://fermi.gsfc.nasa.gov/ssc/data/access/lat/3rd_PSR_catalog/}}. At the position defined by the Galactic longitude $l$ and latitude $b$ of the pulsar, we retrieve $F_{\rm min}$, the minimum detectable flux at that location. If Fermi/LAT did not observe the region, this is also taken into account. A pulsar is considered detectable in $\gamma$-rays if its $\gamma$-ray flux $F_{\gamma}$ exceeds the local threshold $F_{\rm min}$.
%Two criteria must be met for a pulsar to be considered detectable. First, the source must be bright enough. If the pulsar’s Galactic latitude is less than 2°, we adopt a minimum flux threshold of $F_{\rm min} = 4 \times 10^{-15}$ W.m$^{-2}$; for blind searches, this threshold increases to $F_{\rm min} = 16 \times 10^{-15}$ W.m$^{-2}$. If the $\gamma$-ray flux $F_\gamma$ exceeds $F_{\rm min}$, this first condition is satisfied. The second condition relates to the dispersion measure (DM), for which we impose a threshold of 2.5~cm$^{-3}$.pc—consistent with the lowest DM values reported in the Third Fermi Large Area Telescope Catalog of Gamma-ray Pulsars (3PC), as also done in \citet{ghf+18}. If both criteria are fulfilled, the pulsar is considered detectable in $\gamma$ rays.

\section{Results} \label{sec:results}
%\com{Cette section est trop longue. Il faut scinder en sous-sections.}
\subsection{Radio vs $\gamma$-ray pulsars}

\begin{figure*}
    \centering
    \includegraphics[width=0.33\textwidth]{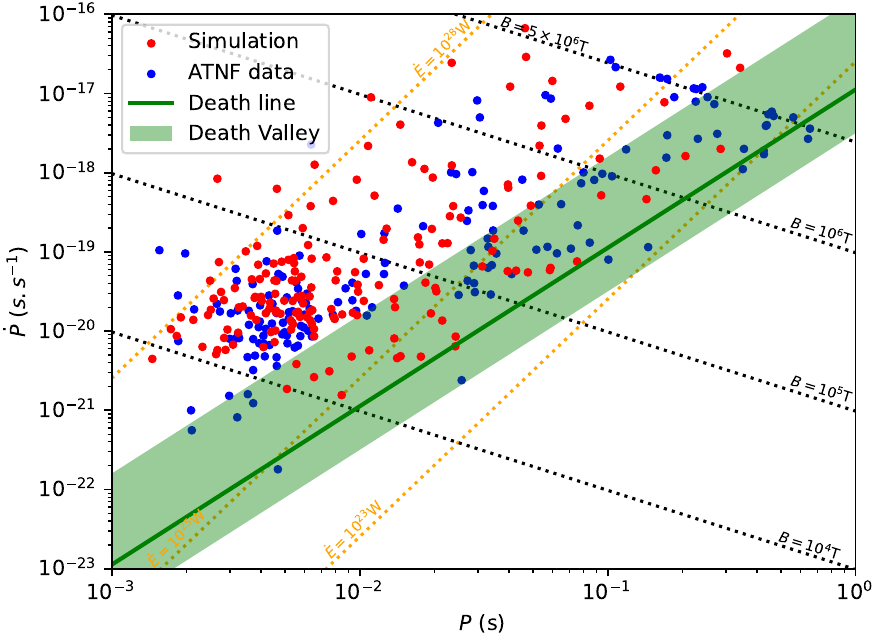}
    \hfill
    \includegraphics[width=0.33\textwidth]{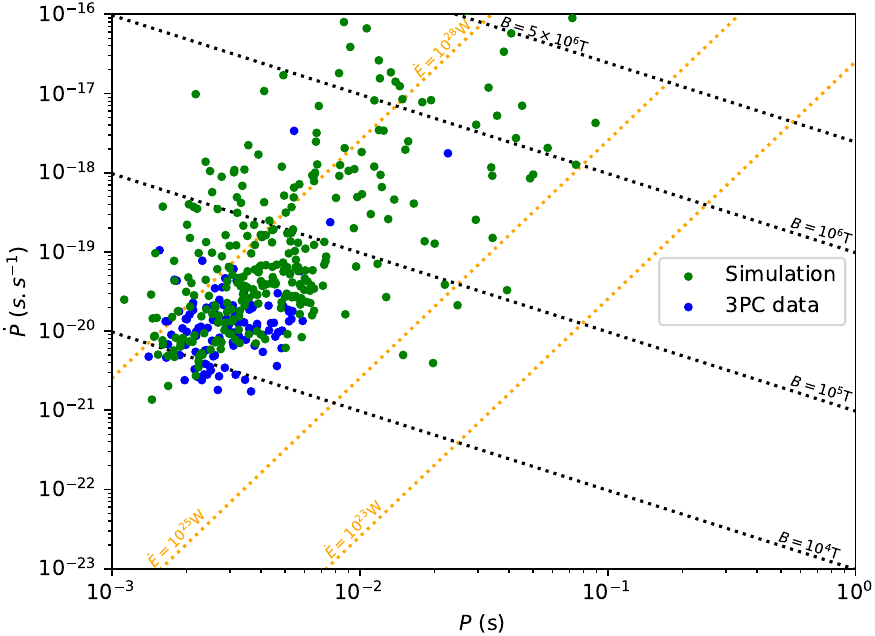}
    \hfill
    \includegraphics[width=0.33\textwidth]{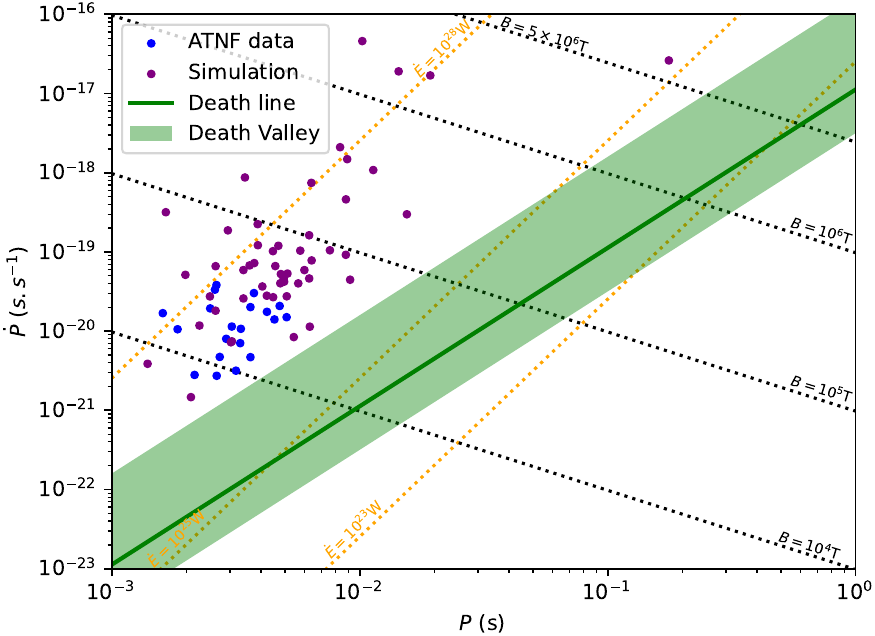}
    \caption{Left panel: $P-\dot{P}$ diagram for the detected pulsars in radio surveys only in the simulations in red, along with the observations in radio for the FAST GPPS or PMPS survey in blue. Middle panel: $P-\dot{P}$ diagram for the detected pulsars in $\gamma$-ray surveys only in the simulations in green, along with the observed pulsars in $\gamma$-ray from Fermi surveys only in blue. Right panel: $P-\dot{P}$ diagram for the simulated pulsars detected both in radio and $\gamma$-ray surveys in purple, along with observations for the FAST GPPS or PMPS survey simultaneously with the $\gamma$-ray observations from Fermi in blue. For all the panels, the death line is represented as a green solid line, and the death valley is the shaded green area.}
    \label{fig:PPdot_rg}
\end{figure*}

The first goal of this study was to match the $P-\dot{P}$ diagram between simulation and observations in order to assess whether the recycling process could primarily explain the observed pulsar population of mildly and fully recycled pulsars in the $P-\dot{P}$ diagram (i.e., pulsars with $B \leq 5\times10^6$~T). By using the parameters for the various distributions at birth and the evolution model presented in Sec.~\ref{sec:model}, we obtained the three $P-\dot{P}$ diagram in Fig.\ref{fig:PPdot_rg}. The left panel displays the diagram for pulsars detected only in radio surveys, the middle panel for those detected only in $\gamma$-rays surveys, and the right panel for pulsars detected in both radio and $\gamma$-rays surveys. In each case, observations are shown alongside simulations. For the radio-only population (left panel), the Kolmogorov–Smirnov (KS) test \citep{s48} yields a p-value of \matteo{$5.6\times10^{-5}$} for the spin period $P$ distribution and \matteo{0.2} for the spin-down rate $\dot{P}$ distribution. This suggests that the null hypothesis is rejected for both $P$ and $\dot{P}$. Nevertheless, the simulated population matches observations well for $P < 0.1$~s, with p-values greater than 0.05 for both $P$ and $\dot{P}$, indicating that few recycled pulsars are observed above this value, meaning this channel of formation only allows to obtain a few mildly recycled pulsars but not all of them. In the ATNF catalog, \matteo{167} radio pulsars detected by the FAST GPPS or PMPS surveys have both measured values of $P$ and $\dot{P}$. Additional detections exist but are sometimes missing a $\dot{P}$ measurement, and were therefore excluded from the comparison with our simulation, or are not yet included in the catalog. A more complete list of such sources can be found online\footnote{\url{https://pages.astro.umd.edu/~eferrara/GalacticMSPs.html}}. In our simulation, we detect \matteo{176} radio pulsars, which agrees reasonably well with the number of radio pulsars discovered by FAST GPPS and PMPS. Furthermore, the death valley eliminates $\sim100$ pulsars, and most of them were well below the valley. Regarding, the middle panel (pulsars in $\gamma$-ray surveys only), most of the observed pulsars appear to be reproduced in the simulation, though there is a noticeable over-detection for $\dot{P} > 3 \times 10^{-20}$. From 3PC, 137 pulsars were selected for comparison (still with the criterion:  $B \leq 5\times10^6$~T), of which 20 were also detected by the selected radio surveys, leaving 117 as pulsars in $\gamma$-ray surveys only. In the simulation, \matteo{98} such pulsars were detected with $\dot{P} < 3 \times 10^{-20}$. When comparing this observed sample to the simulated one (restricted to $\dot{P} < 3 \times 10^{-20}$), the KS test yields p-values of \matteo{0.15} for $\dot{P}$ and \matteo{0.25} for $P$, indicating good agreement. Furthermore, the number of detected pulsars in both datasets is very similar. The overabundance of simulated pulsars with $\dot{P} > 3 \times 10^{-20}$ could correspond to unidentified sources in the Fourth Fermi-LAT catalog of $\gamma$-ray sources (4FGL), or may reflect limitations of the striped wind model. These interpretations are further discussed in Sec.~\ref{sec:discussion}. Finally, the population of pulsars detected in both radio and $\gamma$-ray surveys (right panel) is too small for a robust KS test. However, the simulated pulsars occupy a similar region of the $P$–$\dot{P}$ space as the observed ones. The simulation does show an excess of approximately 20 pulsars compared to observations, which corresponds well to the number of radio MSPs associated with $\gamma$-ray sources that have not yet been observed to pulsate in the 4FGL catalog \citep{aab+22}. These observed sources may simply not have shown detectable pulsations so far, and further investigations could reveal their MSP nature, which would be consistent with our simulation results.
%might also suggest that a few more unidentified pulsars in 4FGL could be detected in radio.

Although the middle panel displays what we classified as pulsars in $\gamma$-ray surveys only, we still computed their expected radio flux using Eq.~\eqref{eq:rad_lum}. This choice is motivated by the fact that, once a pulsar is detected in $\gamma$ rays, it is often followed up with dedicated radio observations. As a result, \matteo{most of the} $\gamma$-ray pulsars eventually have some form of radio measurement associated with them. However, these follow-up observations are not linked to any specific survey. As a result, in the ATNF data, many $\gamma$-ray pulsars that do emit in radio are not listed as being associated with a radio survey, which is the criterion we used to classify pulsars as either radio-loud or radio-quiet in Fig.~\ref{fig:PPdot_rg}. Consequently, $\sim71$\% of the objects we label as pulsars in $\gamma$-ray surveys only in the simulation actually have radio fluxes greater than 30~µJy and a geometry that would allow to see pulsations, i.e $|\zeta-\chi| \leq \rho $, which in real observations would almost always lead to a radio detection \citep{saa+23}. Indeed, in the sample of 117 pulsars used in this work from the 3PC catalog labeled as pulsars in $\gamma$-ray surveys only, only 6 of them actually lack any radio detection, either from surveys or from targeted observations. This corresponds to 94.9\% of $\gamma$-ray pulsars not seen in a radio survey, which have a radio counterpart.%, reinforcing the consistency between our simulation and the observations.

\begin{figure}[h]
\resizebox{\hsize}{!}{\includegraphics{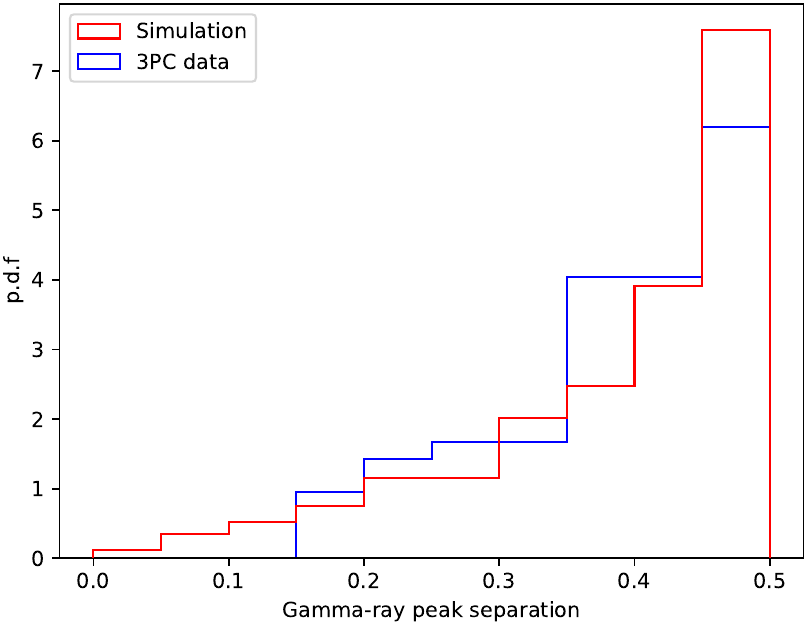}}
\caption{Distribution of the $\gamma$-ray peak separation of the simulated population along with the observations from the 3PC catalogue.}
\label{fig:histo_gpeaksep}
\end{figure}

We can compare the simulated and observed $\gamma$-ray light-curve peak separation $\Delta$ in Fig.~\ref{fig:histo_gpeaksep}. The separation $\Delta$ is computed according to \citet{p11} by
\begin{equation}
\cos\left(\pi \, \Delta\right) = |\cot\zeta \cot\chi | 
,\end{equation}
Here, $\zeta$ is the angle between the line of sight and the rotation axis, and $\chi$ the magnetic inclination angle. All distributions are normalised to yield probability distribution functions (pdfs), by dividing each pdf by the total number of pulsars—observed or simulated. Since the signal is periodic and the definition of the main peak is arbitrary, the peak separation $\Delta$ is restricted to the interval [0, 0.5]; values larger than 0.5 are mapped to $1 - \Delta$ via a phase shift. Figure~\ref{fig:histo_gpeaksep} shows a striking agreement between the observed and simulated $\gamma$-ray peak separation distributions, indicating that the striped wind model reproduces this geometric feature well, as it was also the case for the normal pulsars in \citet{spm+24}. Moreover, for the observed sample, separations smaller than 0.15 are likely unresolved, which explains why this region shows poorer agreement with the observations. Nevertheless, the overall trend of the curve remains consistent between the observations and the simulations.

\begin{figure}[h]
\resizebox{\hsize}{!}{\includegraphics{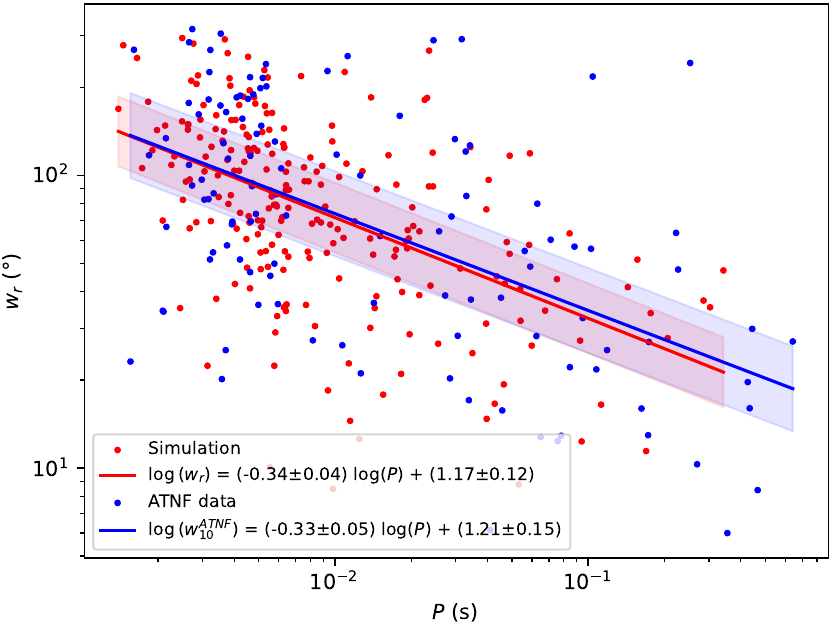}}
\caption{Width of the radio profile plotted against the spin period $P$ of the radio detected pulsars in the simulation along with the observations.}
\label{fig:w_r_plot}
\end{figure}

The similarity in the emission geometry between the simulation and the observations is also evident in Fig.~\ref{fig:w_r_plot}, where we compare the width of the radio profile as a function of the spin period~$P$ for both, simulated pulsars above detection threshold and observed ones. A linear fit in log scale to both datasets yields consistent trends, with $w_r \propto P^{-0.34\pm0.04}$ for the simulation and $w_{10}^{\rm ATNF} \propto P^{-0.33\pm0.05}$ for the observations. The simulation successfully reproduces the full range of observed pulse widths, further indicating that the adopted radio emission geometry is realistic. Although one might expect a larger ratio between the two curves (here $\sim0.97$) if the radio profiles were purely Gaussian, since we compare the full simulated profile width according to eq.~\eqref{eq:observedwidthprofile} with the observed width measured at $10$~\% of the peak intensity, a ratio closer to $\sim1.7$ would be anticipated. Nevertheless, this difference can likely be attributed to the use of a dipolar geometry, to uncertainties in the measurement of $w_{10}^{\rm ATNF}$, and to the skewness of some observed profiles.

\subsection{Companion}

\begin{figure*}
    \centering
    \includegraphics[width=0.48\textwidth]{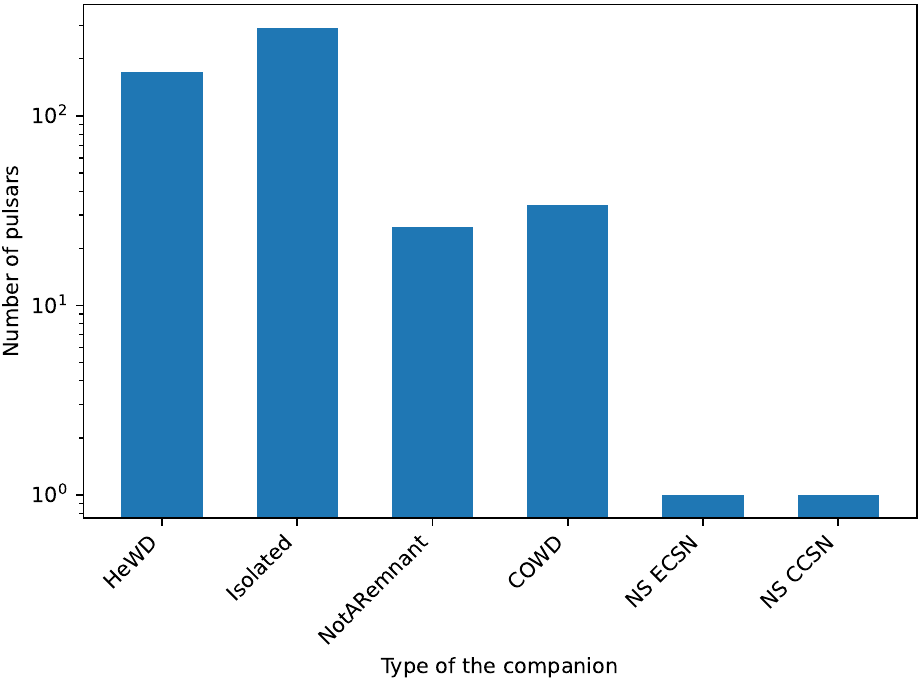}
    \hfill
    \includegraphics[width=0.48\textwidth]{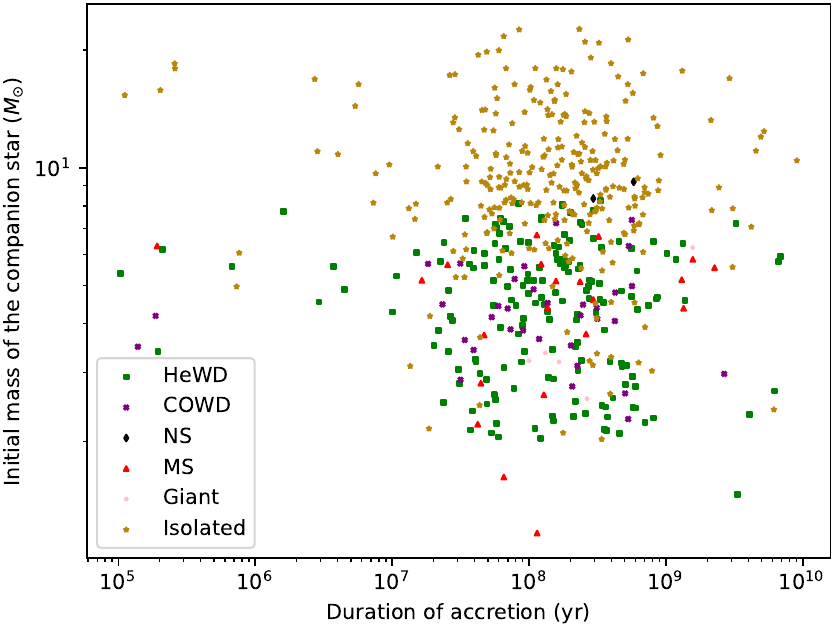}
    \caption{Left panel: Distribution of the type of the pulsars companions. HeWD stands for helium white dwarf, isolated means the pulsar has no companion anymore, COWD stands for carbon-oxygen white dwarf, NotARemnant means the companion is a non-degenerate star, NS CCSN stands for a neutron star which is born after a core collapse supernova, NS ECSN stands for a neutron star which is born after an electron capture supernova and ONeWD stands for oxygen-neon white dwarf. %and BH stands for black hole.
    Right panel: Initial mass of the pulsars companions plotted against the duration of the accretion phase, with the final type of the companion displayed. MS stands for main sequence star and NS stands for neutron star.}
    \label{fig:comp_info}
\end{figure*}

Since every pulsar in the simulation begins with a companion, it is instructive to investigate their eventual fate. The left panel of Fig.~\ref{fig:comp_info} shows the types of companions associated with neutron stars at the end of the simulation (noting that some neutron stars end up isolated). Most of the detected simulated systems ends either as isolated neutron stars or with a helium white dwarf (HeWD) or carbon–oxygen white dwarf (COWD) companion. Only a few systems retain other types of companions, such as non-degenerate stars, other neutron stars, \matteo{oxygen–neon white dwarfs (ONeWDs)} or even a black hole in some cases \matteo{(although this simulation did not produce any BH companion)}. 
%\matteo{(although this simulation did not produce any NS or BH companion)}.
%, oxygen–neon white dwarfs (ONeWDs), or—in one case here—a black hole.
The right panel displays the initial companion mass as a function of the accretion duration. This plot informs us quite clearly that above an initial mass of 8~$M_{\odot}$ for the companion, the survival of the binary is very rare. Most of the pulsars that end up isolated, had a companion whose initial mass was above 8~$M_{\odot}$. Therefore, because of the supernova explosion, most of the time the binary is disrupted by the kick velocity of the companion at the remnant birth. This explains the rarity of NS-NS or NS-BH systems: their formation requires a companion with an initial mass above 8~$M_{\odot}$, followed by a sufficiently low natal kick at the neutron star’s birth to prevent the binary from being disrupted.

A further noticeable aspect of Fig.~\ref{fig:comp_info} is the fact that the initial mass of the companion is not correlated with the duration of the accretion phase of the binary. Basically, most of the binaries have an accretion phase which lasts between $10^8$ and $10^9$~yr. Furthermore, with equation \eqref{eq:diff_eq_model2} we can deduce a typical timescale of alignment for the rotation axis of the pulsar and the normal to the orbital plane, $t_{\rm align}^{\rm orbit}$
\begin{equation} \label{eq:t_align_orbit}
t_{\rm align}^{\rm orbit} = \frac{I \Omega}{n_1},
\end{equation}
where $n_1$ is the accretion torque, $I$ the moment of inertia and $\Omega$ the frequency of rotation of the pulsar. If we compute this timescale with $M_{\rm NS}=1.4 \ M_{\odot}$, $\dot{M}_{\rm NS} = 10^{-9} M_{\odot}$.yr$^{-1}$, $R=12$~km, $B = 5\times10^6$~T, $\Omega = 1$~rad.s$^{-1}$ and $I=10^{38}$~kg.m$^{2}$, which are typical values for a pulsar which have been spinning down for a long time without accretion, it gives $t_{\rm align}^{\rm orbit} \sim 5.9$~kyr. Even if we change these values, in many cases, the typical duration of spin-orbit alignment is going to be lower than the typical duration of accretion, meaning most of the population has its spin-orbit aligned. Indeed, we already showed these results in \citet{lps25} (see Fig.~7), corresponding to $\sim 80$~\% of the population with the spin-orbit angle, $\alpha$, that is less than 10°. This supported the possibility of detecting pulsars with a misaligned spin-orbit (even though simulations show that such cases are relatively rare), as this study specifically investigated the spin-orbit alignment hypothesis. 

%A last remarkable feature in Fig.~\ref{fig:comp_info} is that there can be very massive stars, up to 100~$M_{\odot}$, that allow the neutron star to accrete for $10^8$~yr and above, and that seem odd at first sight as massive stars are not supposed to live for such a long time. Nevertheless, in SEVN, the process of rejuvenation of stars is taken into account. This process, described in \citet{tap+97}, consists of the fact that stars that either gain or lose mass, are going to have their lifetime either decreased or increased. Therefore, the simulation shows that even very massive stars can have their lifetime highly extended thanks to accretion from their primary star in a binary. 

\subsection{Distribution of NS masses}
\begin{figure}[h]
\resizebox{\hsize}{!}{\includegraphics{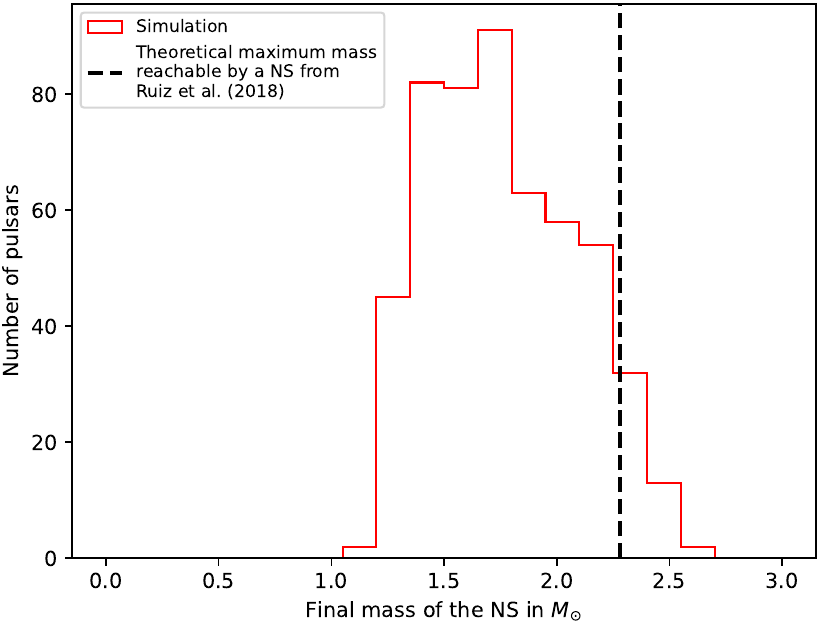}}
\caption{Histogram of the final mass of the recycled pulsars in the simulation. The vertical black dashed line represents the theoretical maximum mass reachable by a neutron star from the study of \citet{rst18}.}
\label{fig:histo_final_mass}
\end{figure}

Studying the mass that neutron stars can reach is very important in order to constrain the equation of state (EoS) \citep{of16}. In this work we are able to follow the mass evolution of the neutron stars, and especially look at what is the distribution of mass characterizing the recycled pulsars detected in the simulation. In Fig.~\ref{fig:histo_final_mass}, the pdf of the detected pulsar mass obtained in the simulation is represented. \citet{l20} gives a review of the neutron star mass distribution observed. An histogram including 44 mass measurements of pulsars was shown (Fig.~4 of \citet{l20}), and at that time it was very new to obtain measurements of pulsars with masses greater than 2~$M_{\odot}$, and for most of them they were spiders system. To this day, the number of measurements of mass of pulsars did not increase much\footnote{\url{https://www3.mpifr-bonn.mpg.de/staff/pfreire/NS_masses.html}}. Nevertheless, many of the observed mass measurements correspond to young pulsars, which clearly did not have time to undergo recycling. Moreover, the total number of available measurements remains limited. In contrast, in this type of simulation, we can obtain the mass of each recycled pulsar. The results suggest that due to accretion, the mass distribution of recycled pulsar is centered around 1.8~$M_{\odot}$, with pulsars potentially reaching up to 2.7~$M_{\odot}$. This is consistent with recent mass measurements of spider pulsars mentioned earlier. Therefore, the simulation indicates that if more masses of recycled neutron stars were measured, we could expect to find significantly more high masses compared to normal pulsars. Although the maximum neutron star mass imposed in SEVN was 3~$M_{\odot}$, no pulsar in the simulation reached this value. There is also the possibility that the pulsars in the simulation accrete too much matter, as we did not observe many high-mass pulsars. Therefore, to diminish the accretion in the simulation, it would require larger $\tau_d$ so that the magnetic field could decay more slowly. This, in turn, would increase the inner disk radius $r_{\rm in}$ for a longer period, thereby shortening the accretion phase. We tested this hypothesis in the simulation, but it failed to reproduce a realistic $P-\dot{P}$ diagram, making this scenario unlikely. Besides, as already mentioned, the number of measured masses of neutron stars is small (< 100). Overall, this study supports the recent high mass measurements of spider pulsars and suggests that the maximum mass of neutron stars may be higher than the 2.28~$M_{\odot}$ proposed by \citet{rst18}, otherwise the model would not allow to reproduce that well the $P-\dot{P}$ diagram.

\subsection{Spatial distribution}
\begin{figure}[h]
\resizebox{\hsize}{!}{\includegraphics{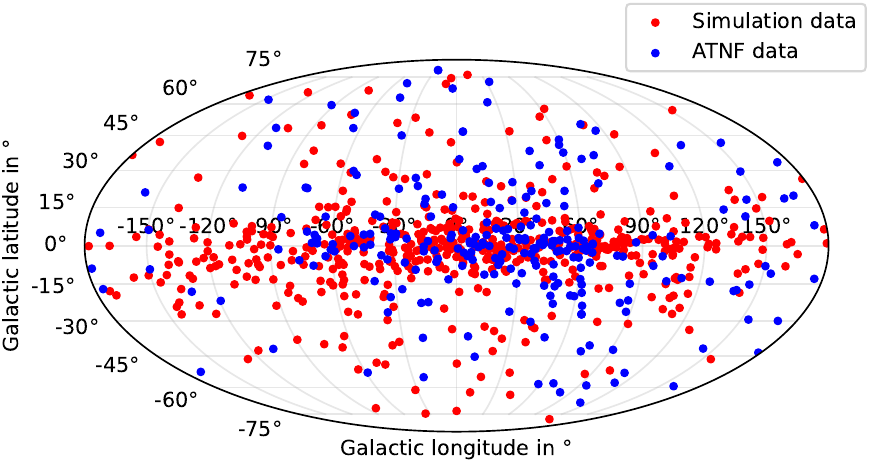}}
\caption{Distribution in Galactic coordinates of the detected recycled pulsars in the simulations along with the observations.}
\label{fig:XY_galacticplane}
\end{figure}

Very interesting result concerns Fig.~\ref{fig:XY_galacticplane}, where we can see the distributions in Galactic coordinates for the detected pulsars in the simulations and in the observations. It is striking that we recover well the positions of the observed pulsars with our simulations. However, we detect especially more pulsars at latitudes between -15° and 15° in the simulation, therefore we could expect many unidentified sources of 4FGL to lie at these latitudes. Furthermore, regarding the measured positions, there are uncertainties which can be greater than 20~\% on their measurement \citep{vwc+12,ivc16,ymw17}, which one must have in mind while comparing the simulated and observed positions of pulsars. In addition, this simulation can give us an estimation of the number of recycled pulsars born in the Galactic field which end up in the Galactic center. Two majors hypothesis exist about the origin of the GeV excess, either it is a self-annihilating dark matter particles origin \citep{ccw15,aaa+17,h22}, either it is of pulsars origin \citep{ya16,m24,spm+24}. Moreover \citet{ya16} computed a total luminosity of $10^{30}$~W on the inner 10°$\times$ 10° region of the Galactic center, which corresponds to a projected square of 1.4~kpc $\times$ 1.4~kpc around the Galactic center (assuming a distance of 8.5~kpc from the Sun), i.e., a radial extent of $\sim$ 0.7~kpc. In the simulations, it would correspond to a total of $\sim4.6\times10^{28}$~W generated by $\sim190$ $\gamma$-ray pulsars on average in a 0.7~kpc radius from the Galactic center (whether these pulsars are detected or not, and usually there is between 0 and 2 $\gamma$-ray detection in that area in the simulations). Therefore, the simulation hint to the fact that MSPs that were originally born in the Galactic field seem to contribute to $\sim5$~\% of the GeV excess.
%The whole sample of pulsars have luminosities between $8.5\times10^{24}$~W and $1.0\times10^{29}$~W with a mean luminosity of $1.5\times10^{27}$~W, which means there should be between $10$ and $120,000$ $\gamma$-ray pulsars from the Galactic field in the Galactic center to explain the GeV excess, with a more likely value of $\sim700$ pulsars in the Galactic center.
In addition, \citet{spm+24} also found that young $\gamma$-ray pulsars could contribute to this excess with their detection prospects, and we discuss also the detection prospects of the recycled pulsars in Sec.~\ref{sec:discussion}.

\begin{figure}[h]
\resizebox{\hsize}{!}{\includegraphics{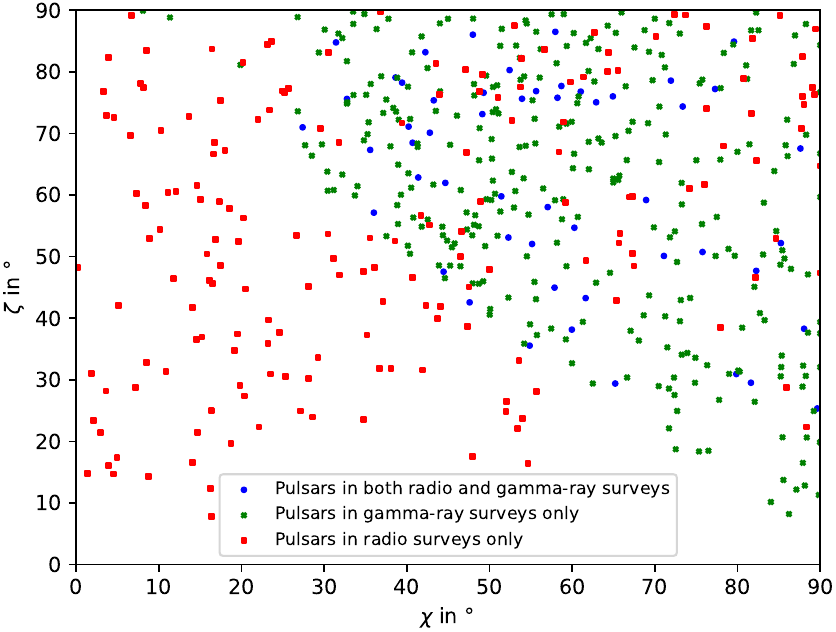}}
\caption{The viewing angle $\zeta$ plotted against the inclination angle $\chi$ for the detected pulsars in the simulation.}
\label{fig:zeta_chi_all}
\end{figure}

\subsection{Geometry}
Examining the geometry of the pulsars detected in various wavelengths is interesting. Indeed, Fig.~\ref{fig:zeta_chi_all} shows the viewing angle, $\zeta$, as a function of the inclination angle, $\chi$ for each detected pulsar in the simulation. Although these angles are generated between 0 and 180°, we folded them between 0 and 90°. Basically, we retrieve an area of detection slightly different for radio pulsars compared to \citet{jsk+20} where they studied the normal pulsars, but almost the same area of detection for $\gamma$-ray pulsars. The radio pulsars are detected along a region much broader than along the $\zeta=\chi$ diagonal as it was the case for normal pulsars, the pulsars in $\gamma$-ray surveys only are seen above the line of equation $\zeta=90 - \chi$, and the pulsars detected in both radio and $\gamma$-ray surveys are the intersection of these two areas. The area of radio detection is broader for recycled pulsars compared to normal pulsars, because recycled pulsars have larger $\rho$ in general, which broaden the area of detection $|\zeta - \chi| \leq \rho$ (or $|\zeta - (\pi-\chi)| \leq \rho$). We remind that, $\sim71$ \% of the pulsars in $\gamma$-ray surveys only have a radio flux above 30~µJy and a geometry favorable for a radio detection, thus many of them would still be seen with a follow-up observation in radio. One can notice that it seems particularly difficult to see a pulsar with a rotation axis and magnetic axis aligned, i.e., $\chi \approx0°$. Indeed, for $\gamma$-ray detection the condition $|\zeta - \pi/2| \leq \chi$ is needed, and below $\chi=20$° it almost never happens. The rarity of low detected $\chi$ values appears to be due to the fact that, as shown in \citet{yl23}, the effect of accretion tends to lead to a misalignment between the rotation axis and the magnetic axis, i.e., having $\chi$ increasing towards 90°. Therefore, after accretion, since the magnetic field is weaker, the torque $n_3$ caused by loss of energy via radiation of the pulsar has more difficulties to get $\chi$ to decrease. Furthermore, it is also plausible for those which get low $\chi$ that it took too much time and they became not luminous enough to be seen. These findings on the simulated population reflect well the observed values for $\zeta$ and $\chi$. For instance \citet{bpm21} constrained these angles for 10 radio-loud $\gamma$-ray MSPs, and found out that for both angles they are usually above 45°, and it is true that in our simulation the radio-loud $\gamma$-ray pulsars also verify this. Another example is the study of \citet{gt14}, where they concluded that $\gamma$-ray energetic nearby MSPs are not seen because of their unfavorable viewing conditions, i.e., low $\zeta$ and we also see that high $\zeta$ are more favorable conditions of observations. 

\begin{figure*}
    \centering
    \includegraphics[width=0.49\textwidth]{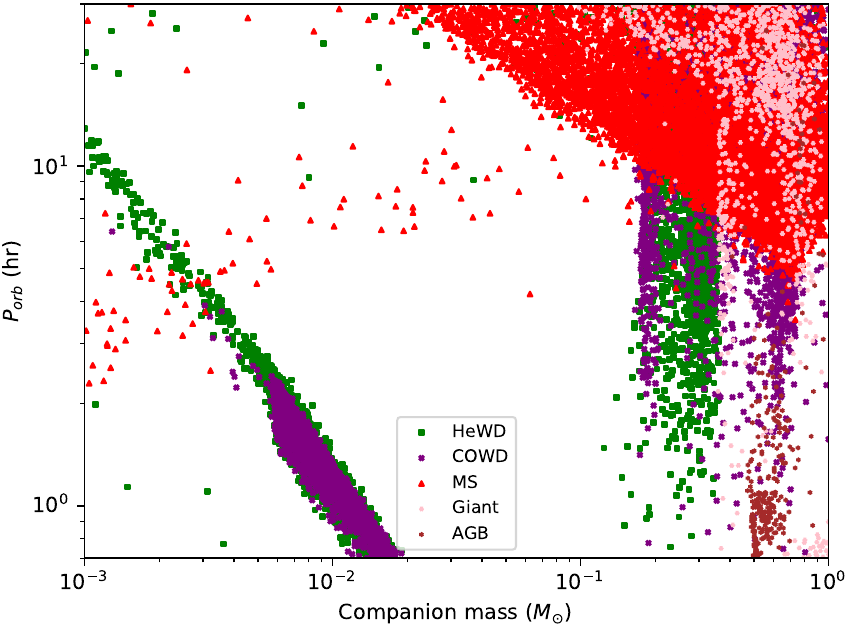}
    \hfill
    \includegraphics[width=0.49\textwidth]{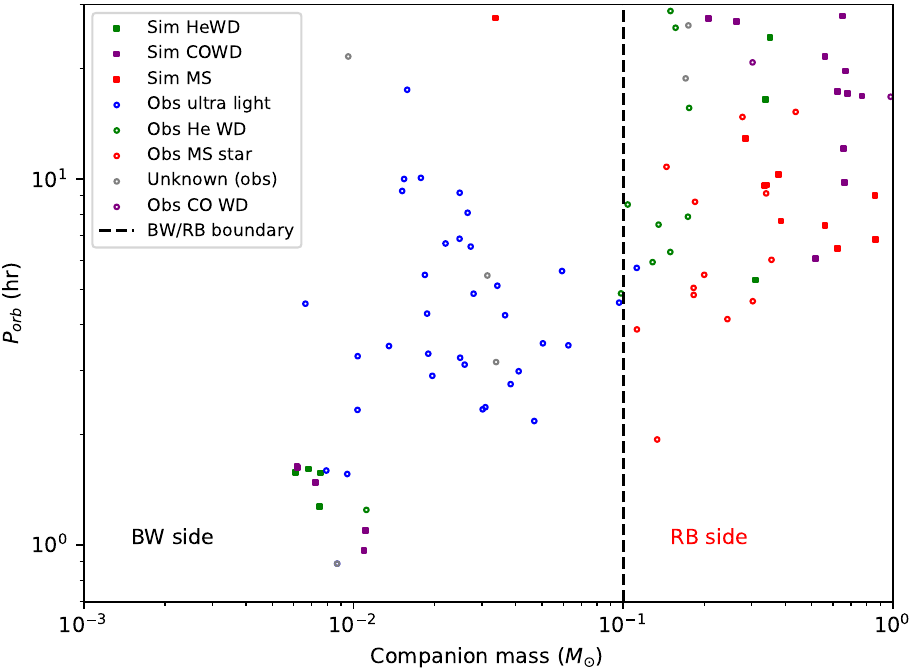}
    \hfill
    \caption{Orbital period, $P_{\rm orb}$ (in hr), as a function of companion mass (in $M_{\odot}$). AGB stands for a companion that is on the asymptotic giant branch. Left panel: entire simulated population. Right panel: Only systems with a detected pulsar by the pipeline in the simulation and observed systems. The companion mass reported is the median mass available in the ATNF catalog for the observed systems. Unknown (obs) means it has a null value in the ATNF catalog, "Sim" stands for simulated, "Obs" stands for observed. The simulated systems are represented by squares and observed systems by circles, the different colors are for the different companions and are shown in the legend. }
    %\caption{Orbital period, $P_{\rm orb}$ (in hr), as a function of companion mass (in $M_{\odot}$). Left panel: entire simulated population. Middle panel: only systems with a detected pulsar by the pipeline. Right panel: Observed systems. The companion mass reported is the median mass available in the ATNF catalog. Unknown companion means it has a null value in the ATNF catalog.}
    \label{fig:P_orb_comp_mass}
\end{figure*}

\subsection{Spider population}

We may ask whether, at the end of the simulation, it is possible to identify spider pulsars within the recycled population. \matteo{A key process governing the evolution of spider systems is the irradiation of the donor star by the pulsar wind as demonstrated in the works of \citet{cct+13,bdh14,mly25}. In particular, companions with masses between $0.1$ and $1 \ M_{\odot}$ can lose a significant fraction of their mass, causing the orbital period to first decrease and then increase again after a few Gyr (see Fig.~3 of \citet{bdh14}, Figs.~1–4 of \citet{cct+13}, and Fig.~1 of \citet{mly25}). We model this effect using the following equation 
\begin{equation} \label{eq:irradiation_eq}
\dot{M}_{\rm c,evap} = - f_{\rm evap}\frac{L_{\rm sp}}{2 \ v^2_{\rm c,esc}} \left(\frac{R_{\rm c}}{a}\right)^2 .
\end{equation}
Indeed, the irradiation allows a mass loss $\dot{M}_{\rm c,evap}$ from the companion, which depends on the escape velocity from the companion star surface $v_{\rm c,esc}$, its radius $R_{\rm c}$, the semi-major axis of the binary $a$, the pulsar spin-down luminosity $L_{\rm sp}$ and the efficiency parameter $f_{\rm evap}$ measuring how effectively the pulsar's spin-down luminosity is converted into an evaporative wind. For our simulations we set $f_{\rm evap}=0.05$, as having greater values seem to make difficult the production of black widows systems \citep{cct+13}.
}A typical diagnostic to distinguish redbacks (RBs) from black widows (BWs) is a plot of the binary orbital period ($P_{\rm orb}$) versus the pulsar's companion mass in $M_{\odot}$\matteo{, as in the right panel of Fig.~\ref{fig:P_orb_comp_mass} with the observed systems} (see also Fig.~1 of \citet{bgv+25}). In Fig.~\ref{fig:P_orb_comp_mass}, the left panel shows this diagram for the entire simulated population, while the right panel presents the same plot restricted to the detected systems \matteo{in the simulation and observed systems}. %Strikingly, \matteo{the right panel} reveal the presence of two distinct groups \matteo{in the simulation}: 
RBs are located in the upper right, and BWs appear in the left region of the plot. To compute $P_{\rm orb}$ in the simulation, we used Kepler's third law which give
\begin{equation} \label{eq:kepler_3rd_law}
    P_{\rm orb} = \sqrt{\frac{4 \pi^2 a^3}{G \left(M_{\rm c} + M_{\rm NS}\right)}} \ ,
\end{equation}
where $M_{\rm c}$ is the companion's mass, $M_{\rm NS}$ is the neutron star mass, $a$ is the semi major axis of the binary and $G$ is the gravitational constant. As a consequence, in a log-log plot of $P_{\rm orb}$ against $M_{\rm c}$, we can actually observe a line with a $-1/2$ slope in \matteo{the left} panel for the BWs as they also have a close orbit, thus the dependence in $a$ is diminished, in the contrary to the RBs which are in wider orbits.
Observed RBs typically have main sequence companions\matteo{, we can observe 13 systems like this}, whereas our simulation \matteo{produced 9 of them here.} %\matteo{in some realization of the simulation (and none on this example)}.
We find \matteo{slightly more} pulsars \matteo{in the RB region} with carbon-oxygen white dwarf (COWD) companions \matteo{compared to the observations}. Observed BWs have ultra-light companions, \matteo{while} most simulated systems in the BW region host either HeWDs or COWDs, \matteo{suggesting that some of the companion of observed BWs could actually be \matteo{HeWDs or} COWDs, or we could discover such spiders. We also struggle to reproduce BWs systems with a $P_{\rm orb}>2$~hr, whereas, on the left panel, we can see that a few systems with the characteristics of BWs with $P_{\rm orb}>2$~hr are formed but not detected. In addition, there are currently around 32 confirmed RBs and 50 confirmed BWs \citep{kl25}, whereas our simulation produces roughly three times fewer spider-like pulsars. These discrepancies may indicate that a refinement is necessary for the initial conditions concerning the semi-major axis $a$ which determine the initial orbital period, or combining a different set of parameters for the efficiency parameter of irradiation $f_{\rm evap}$ and the accretion efficiency $f_{\rm MT}$ or even other parameters describing the binary evolution. It may also hints that these systems might often form through alternative channels, as proposed by \citet{mw25}. These authors suggest that the ultra-light companions of spiders could form in high-metallicity accretion disks over timescales exceeding $> 1$~Gyr \citep{pk17}, through dynamical capture or exchange interactions in clusters \citep{srh+03}, or, especially for BWs, by the accretion of an ultra-compact X-ray binary companion’s outer layers onto the neutron star \citep{bbb+11}.} The study by \citet{mly25}, which uses the \matteo{detailed stellar evolution} code MESA to characterize the spider population, shows that an accretion efficiency of 0.7 works better to reproduce the observed systems. \matteo{We did not find any major differences in terms of detections between a simulation with $f_{\rm MT}=0.7$ or also by changing $f_{\rm evap}=0.5$}. In summary, the simulation successfully produces two distinct groups reminiscent of spiders, which is encouraging. However, it underestimates the total number of observable spider systems, therefore including the additional formation channels mentioned above \matteo{and fine-tuning the parameters related to binary evolution} would likely lead to a better agreement with observed populations.

\begin{figure*}
    \centering
    \includegraphics[width=0.48\textwidth]{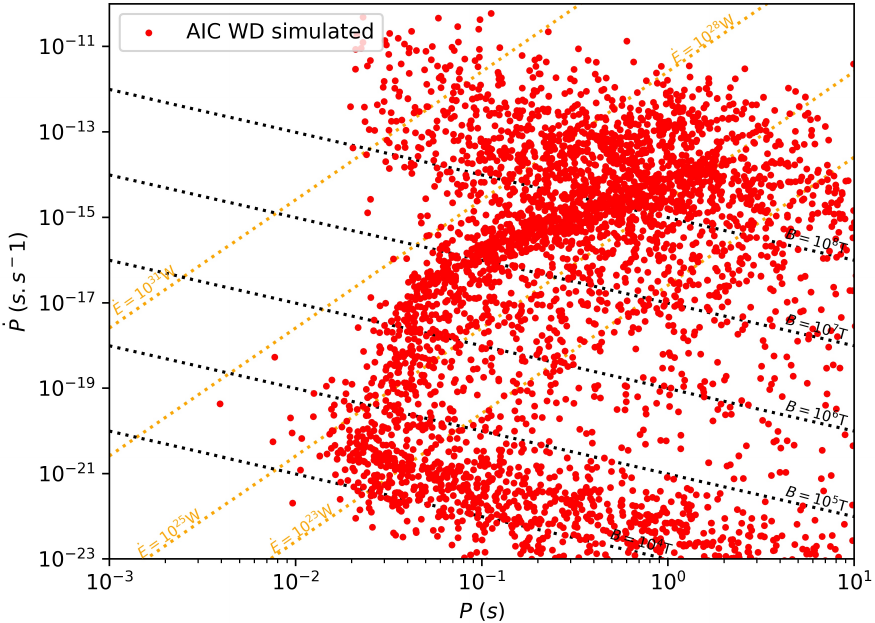}
    \hfill
    \includegraphics[width=0.48\textwidth]{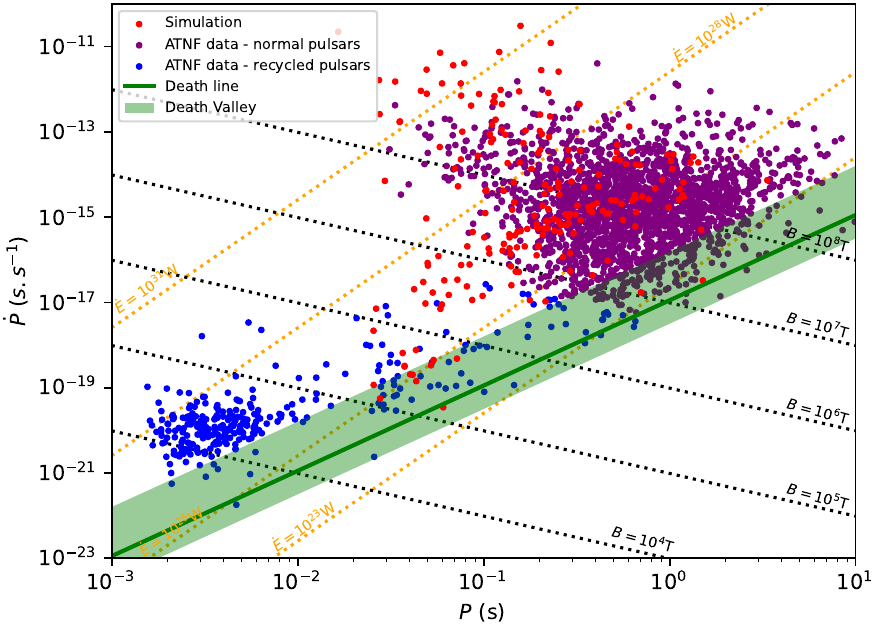}
    \caption{$P-\dot{P}$ diagram of pulsars born after accretion induced collapse (AIC) of white dwarfs. Left panel: entire simulated population. Right panel: only systems with a detected pulsar by the pipeline along with the observations.}
    \label{fig:P_PdotAIC}
\end{figure*}

\subsection{AIC scenario}

Another formation channel for pulsars is the accretion-induced collapse (AIC) of white dwarfs (WDs) \citep{wg23}. We performed a simulation starting with one million binary systems composed of an oxygen-neon (ONe) WD and a main sequence companion, with the latter at a random evolutionary stage (0–100\% of its main sequence lifetime), as was done for the pulsars in this study. We considered only ONe WDs, as their higher mass and composition make them ideal candidates for electron-capture supernovae (ECSNe). Consequently, most of the evolutionary channels leading to AIC involve an ONe WD \citep{wl20}. Additional simulations were conducted with an initial population consisting of one third of each WD type, and only the ONeWDs were found to undergo collapse into neutron stars. The initial mass distributions for the WDs were taken from \citet{tck+16}. Figure~\ref{fig:P_PdotAIC} shows the resulting $P$–$\dot{P}$ diagram: the left panel displays the entire population of neutron stars formed via AIC, while the right panel presents only those that are detectable (either in radio or $\gamma$-rays) based on our detection pipeline, alongside the observed pulsar population. Out of the one million initial systems, \matteo{$\sim 9,000$} WDs evolved into neutron stars (\matteo{$\sim$~0.9\%}). Within the total population, we distinguish two main groups: one consists of neutron stars that accreted little mass and remain relatively young, behaving like "normal" pulsars; the other group lies in the region of mildly recycled pulsars (those with $30 \leq P \leq 1000 $~ms and $\dot{P} < 10^{-16}$~s/s), having accreted significantly more mass and consequently spun up to shorter periods, although it was not sufficient for them to become fully recycled as MSPs. After applying the detection pipeline, only \matteo{$\sim 600$} pulsars remain. Most of these belong to the normal pulsar population. A few objects ($\sim 70$) are found in the mildly recycled region of the $P$–$\dot{P}$ diagram, but they exhibit relatively high spin periods, with none below $20$~ms. This suggests that a non-negligible fraction of recycled pulsars with relatively high spin periods may originate from the AIC of ONeWDs (depending on the birth rates of these objects). As shown in Fig.~\ref{fig:PPdot_rg}, these long-period/mildly recycled pulsars are not fully accounted for by the formation channel involving a neutron star formed via core-collapse supernova (CCSN) that later accretes mass from a companion. This discrepancy explains the imperfect agreement of our KS test for the radio population. The existence of this alternative formation pathway could help alleviate the so-called "birth rate problem" \citep{ge24}, in which the Galactic CCSN rate appears to be lower than the combined estimated birth rates of neutron stars. In summary, this informs that among the whole ONeWDs simulated, \matteo{$\sim$~0.9\%} collapse into neutron stars, \matteo{$\sim$~6.9\%} are detectable and then \matteo{$\sim$~12.2\%} are in the area of mildly recycled pulsars considered in this work. Therefore, \matteo{$\sim 7.6\times10^{-3}$~\%} of ONeWD (that were in a binary system) seem to contribute to the population of mildly recycled pulsars. The contribution of ONeWDs to the pulsar population could be explored in depth in future work, as it would require knowledge of the birth rate of ONeWDs, a quantity that remains highly uncertain.

\begin{figure}[h]
\resizebox{\hsize}{!}{\includegraphics{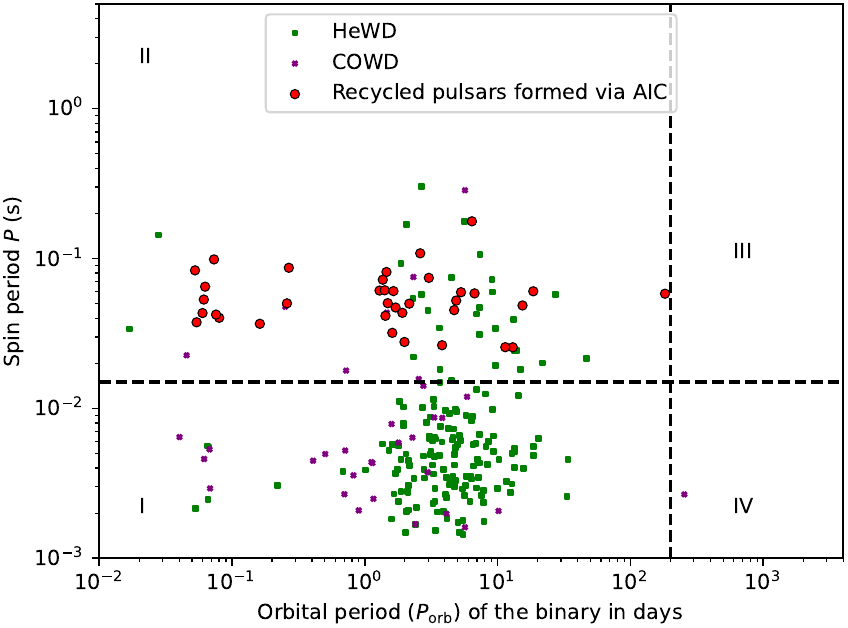}}
\caption{Spin period $P$ plotted against the orbital period $P_{\rm orb}$ of the simulated pulsars. The red circles with black outlines dots, are the AIC WD that became recycled pulsars, while the other points represents recycled pulsars formed via core-collapse supernova with either a He WD or CO WD companion. The four regions, labeled I, II, III and IV represent the same cut as in \citet{tsy+13} to explain their differences as an evolutionary point of view.}
\label{fig:Porb_AIC}
\end{figure}

\matteo{It was suggested in \citet{tsy+13} that the observed pulsars located in region~II of the Corbet diagram (Fig.~\ref{fig:Porb_AIC}) may originate from AIC. In Fig.~\ref{fig:Porb_AIC}, the recycled pulsars originating from an AIC channel are shown as red circles with black outlines dots, whereas the other points correspond to recycled pulsars formed through core-collapse supernovae, paired either with a He~WD or a CO~WD companion. In our simulation, all pulsars formed through AIC indeed end up in region~II. However, we also produce pulsars in region~II that do not have an AIC origin. Therefore, while identifying an observed system as AIC-formed would imply that it should lie in region~II, the reverse is not true: a system found in region~II does not necessarily result from AIC. That said, if a system has $P_{\rm orb} < 1$~day and a $P>10$~ms, an AIC origin appears significantly more likely. }

\section{Discussion} \label{sec:discussion}

\subsection{Model evaluation and caveats} \label{subsec:model_criticism}

The model used seems to reproduce the recycled pulsars in the radio and $\gamma$-ray wavelengths, however it is important to address several points. A crucial aspect of the model is the mass transfer, changing the accretion efficiency, $f_{\rm MT}$, was not explored in this work and we decided to keep the value proposed by SEVN and used in other similar works \citep{bms+21,imc+23}. Nevertheless, changing this value would clearly impose to change some parameters such as, $\tau_d$ and the birth rate for instance.

Furthermore, SEVN does not allow spin-up during the CE phase. Therefore, this is a channel of formation of recycled pulsars from which we do not evaluate the contribution. However, \citet{obg+11} estimated that the accreted matter during the CE phase would be too chaotic to produce spin-up and it could only decay the magnetic field. We also do not take into account the possible accretion from winds. \citet{khb+08} estimated that the spin-up caused by wind would also be too chaotic to produce any significant spin-up. 

\matteo{Still regarding the CE phase, major uncertainties remain about its treatment in binary evolution \citep{th23}, particularly for the formation of double neutron star (DNS) or neutron star–black–hole (NS+BH) systems. When the donor star fills its Roche lobe, a CE phase is triggered with the neutron star, leading to a strong reduction of the orbital separation before the donor undergoes its supernova explosion. Moreover, \citet{ktl+16} showed that an in-spiraling NS can in principle eject the envelope of its companion, but that predicting the final post-CE separation remains highly uncertain. As long as dedicated studies do not clarify the CE evolution for such systems, accurately reproducing the number of observed DNS and NS+BH will remain challenging. In our simulations, we form about $\sim 2- 5$ DNS or NS+BH systems, while observations report 12 DNS systems. Despite these uncertainties, our models are still able to produce a fraction of the observed population, however, the small number of systems involved implies that the statistical significance remains limited.}

The parameter $\epsilon$, characterizing the ellipticity of the neutron star could also be an interesting parameter to vary in order to explore the possibility of formation of sub-millisecond pulsars \citep{hz89}. In this work we only try to draw this parameter from a uniform distribution between the minimum and maximum value of $\epsilon$ hypothesized by \citet{gc22}. The works of \citet{b98,cmm+03} suggested indeed that GW radiation allowed to keep the spin period $P$ above the millisecond threshold, therefore by giving the same $\epsilon$ to every neutron star in the simulation and try to find the value to obtain sub-millisecond pulsar could be quite interesting. Everything that is mentioned above could be explored in future works. 

\matteo{Part of the angular momentum of the binary is lost through magnetic braking, and one as to know that we used the formula proposed by \citet{rvj83} for this phenomenon as it is implemented by default in SEVN. However, recently \citet{cth+21} showed that using the prescription of \citet{vih19} of magnetic braking could, among other things, influence a lot the number of NS + WD systems. Thus, in the future, it would also be interesting to evaluate the influence on the population while modifying the magnetic braking model of the binary.}

For the simulation involving ONeWDs and main sequence stars, we adopted the same initial distributions for eccentricity ($e$) and mass ratio ($q$) as used for massive binaries, as described in Sect.~\ref{sec:model}. We also explored alternative power-law indices for the $q$ distribution, varying it between $-0.5$ and $0.5$ instead of the default $-0.1$, but found no significant impact on the simulation outcomes. Similarly, due to the lack of reliable constraints on the birth eccentricity distribution for low-mass binaries, we tested different power-law indices for $e$ as well, and again observed negligible differences in the results. 

Let us mention several geometrical or physical limitations concerning the striped wind model, which may affect our results. For instance, as suggested by \citet{kwh+22}, a more accurate description of the $\gamma$-ray luminosity might involve an additional parameter, such as the cut-off energy, and a different dependence on $B$ and $P$, or equivalently on $P$ and $\dot{P}$, than the one assumed in \eqref{eq:gamma_lum}. Moreover, plasma recollimation effects within the striped wind could reduce the $\gamma$-ray beam opening angle compared to the adopted value, leading to a smaller number of detectable pulsars.

\subsection{Comparison with other works} \label{subsec:comp_other_works}

Recently, \citet{tl25} investigated the populations of both normal and millisecond pulsars in the radio and $\gamma$-ray bands. Their study focused in particular on testing various $\gamma$-ray luminosity models, leading them to rule out three of them—namely, the pair-starved polar cap (PSPC), the slot gap two-pole caustic (TPC), and the outer gap (OG) models—because they overproduced detectable pulsars. However, their overall modeling framework differs significantly from ours: they do not simulate the full accretion process, they do not include magnetic field decay, and they assume an exponential decay of the inclination angle $\chi$ (referred to as $\alpha$ in their work). Moreover, when they adopt the same $\gamma$-ray luminosity prescription as in this study, they do not incorporate the striped wind geometry. As a result, their model predicts a much larger number of detectable $\gamma$-ray sources, ranging from 350 up to 650. In contrast, our more comprehensive approach, which includes magnetic field decay, accretion history, and the striped wind geometry, yields a tighter prediction, with between \matteo{290 and 330} detectable $\gamma$-ray recycled pulsars, reflecting stochastic variations across simulation realizations. This results in fewer than \matteo{$\sim220$} unidentified sources in the 4FGL catalog, significantly narrowing the allowed range compared to the upper limit of $\sim$520 unidentified sources predicted by \citet{tl25}.

\citet{ghf+18} estimated that approximately 11,000 MSPs would be required within the inner Galactic center (defined as the 7°~$\times$~7° region of interest, corresponding to a radius of about 0.5~kpc) to account for the GeV excess. Their scenario relies on the contribution of globular clusters that have migrated into the Galactic bulge through dynamical friction \citep{got14}. In contrast, our study considers only pulsars formed in the Galactic spiral arms. These tend to be more luminous than those assumed by \citet{ghf+18}. Nevertheless, our results indicate that only about 190~MSPs born in the spiral arms contribute to roughly 5~\% of the GeV excess. Therefore, if the GeV excess is entirely due to pulsars, it is still entirely plausible that a much larger population of MSPs originating from globular clusters resides in the Galactic center as estimated by \citet{ghf+18} or they were born in the bulge, but they are beyond current detectability.
%making it even more plausible that a smaller number of such recycled pulsars than their estimation could explain the observed GeV excess. Most of them would remain undetected as individual pulsars due to Fermi-LAT sensitivity limits, which is consistent with the fact that our pipeline identifies only 2 $\gamma$-ray pulsars in this region, but they would indeed contribute to the diffuse $\gamma$-ray emission in that region.

The studies of \citet{khb+08,kh09} were among the first to attempt reproducing the pulsar population by accounting for the full evolutionary path of neutron stars in binary systems. However, several key differences exist between their approach and ours. First, their magnetic field decay prescription is exponential; thus, if they had used the same decay timescales $\tau_d$ as those identified in our work (0.05, 1.4, or 2.3~Myr), the magnetic fields of their neutron stars would decay unrealistically fast. As a result, they require significantly higher values of $\tau_d$ to reproduce the observed population. Secondly, the equation they adopt for the evolution of the angular velocity $\Omega$ is much simpler than the one used in this study, which incorporates several additional physical effects. Furthermore, while our initial distributions for $P_0$ and $B_0$ are informed by observational constraints and recent population synthesis efforts \citep{ifg+22,spm+24}, their work relies on uniform distributions for these parameters, not grounded in observational data or detailed modeling. Finally, their comparison with observations is limited, as their population synthesis framework does not include a detection model to account for selection effects.

As mentioned in Sect.~\ref{sec:model}, we obtain matching $P-\dot{P}$ diagram between observations and simulations for a number of simulated binaries between \matteo{$4.6\times10^5$ and $5.7\times10^5$ corresponding to a birth rate between  $\rm 3.3-4.1 \times10^{-6} \ yr^{-1}$}. Studies by \citet{sgh07} and \citet{ghf+18} found the birth rate of MSPs to be in the range $\rm 3-6.5 \times10^{-6} \ yr^{-1}$. The birth rate found in our study is consistent with the findings of \citet{sgh07} and \citet{ghf+18}, we constrain the birth rate of recycled pulsars even more with this work. 

\subsection{Detection prospects} \label{subsec:detec_prospects}

We wanted to assess the detectability of new recycled pulsars sources with this model. To do so, we considered the survey parameters of the Square Kilometer Array (SKA), see Appendix~\ref{appendix:AppD} to see the parameters, in order to make predictions for radio observations. Moreover, concerning the $\gamma$-ray wavelengths, we took into account an instrument that would be $10$ times more sensitive than Fermi/LAT (i.e., we adopt a detection threshold of $F_{\rm lim} = 1 \times 10^{-16}$~W.m$^{-2}$, as the average flux limit of $\sim 1 \times 10^{-15}$~W.m$^{-2}$ was derived from the Fermi/LAT all-sky sensitivity map). These considerations allow to multiply the number of radio detections by $\sim3$ and the number of $\gamma$-ray detections by $\sim5$. This improvement in sensitivity also allows to detect $\sim10$ $\gamma$-ray pulsars in the Galactic center, indicating that an even higher sensitivity would be necessary to uncover additional pulsars in the Galactic center.
%with a total luminosity of $1.8\times10^{29}$~W, comforting our estimation of having between 400 and 3000 pulsars to explain the GeV excess. 

\section{Conclusions} \label{sec:conclusions}

This study is the first ever to evolve the population of pulsars in binary, in a complete way by taking into account the binary processes, and that also takes into account observational biases from radio and $\gamma$-ray surveys in order to analyze the recycled pulsar population. 

We showed that our model allowed to reproduce most of the recycled pulsars population in the radio and $\gamma$-ray wavelengths, and that the mildly recycled pulsars (those with $30 \leq P \leq 1000 $~ms and $\dot{P} < 10^{-16}$~s/s) seem to originate from AIC of WDs for most of them. In order to reproduce the population, the characteristic B-field decay timescale, $\tau_d$, was chosen to be either 0.05, 1.4 or 2.3~Myr, with most of the population evolving with $\tau_d=0.05$~Myr. Moreover, we estimated with this model a birth rate for recycled pulsars between \matteo{(3.3 - 4.1)~$\times10^{-6}$~yr$^{-1}$}. The striped wind model, allows to reproduce very well the gamma-ray peak separation observed for $\gamma$-ray recycled pulsars. Most of the pulsars accrete for $10^{8-9}$~yr, no matter the type or initial mass of the companion. Since the timescale for alignment between the rotation axis of the neutron star and the normal of the orbital plane is often short, we obtain $\sim$80~\% of the population which has these two axis aligned \citep{lps25}. Moreover, the mean mass obtained for this population is about 1.8~$M_{\odot}$, with some very high mass pulsars, up to 2.7~$M_{\odot}$, totally in agreement with the recent measurements of high masses of pulsars in spider systems for instance \citep{l20}. Besides, we are able to recover in the simulation pulsars that have the characteristics of spiders, but not the whole observed population, pointing out that other channels of formation (in addition to the one considered in this work) must be at work to obtain spiders. Moreover, to obtain a more similar population of spiders, as suggested by \citet{mly25}, it is likely necessary for a fraction of the population to have higher accretion efficiency. The positions of the detected recycled pulsars in the simulation are generally well reproduced, although the simulation predicts more pulsars between latitudes -15° and 15° than observed.
Additionally, we estimate that MSPs formed in the Galactic field contribute only to $\sim5$~\% of the GeV excess, corresponding to roughly 190 MSPs in the Galactic center. This result suggests that if the GeV excess originates only from pulsars, the majority of MSPs in the Galactic center likely come from globular clusters that migrated inward through dynamical friction or they were born in the bulge, either way we are not able to detect them nowadays \citep{got14,ghf+18}.
%In addition, we estimated the $\gamma$-ray luminosity of the detected pulsars in that wavelength, there was a mean luminosity of $1.5\times10^{27}$~W, knowing the total GeV excess measured, we derived an estimation of 10 to 120,000 $\gamma$-ray pulsars that would be necessary to explain the GeV excess, with a more likely number of $\sim700$.
Finally, the detected pulsars seem to have, for most of them, an inclination angle, $\chi$, greater than 10°, and it also seems easier to observe pulsars that have a viewing angle, $\zeta$, greater than 45°, which was already supported by other studies \citep{bpm21,gt14}. We were also able to make some predictions: we expect $\sim3$ times more radio recycled pulsars thanks to SKA and $\sim5$ times more detection of $\gamma$-ray recycled pulsars with an instrument which has a sensitivity increased by one order of magnitude compared to Fermi/LAT. 

%We believe this work could lead to many other works, as it considers the evolution of neutron stars in a binary with many parameters, and many constraints could be found about these parameters as already discussed in Sect.~\ref{sec:discussion}. Furthermore, 
Adding other channel of formation for recycled pulsars, studying in depth the contribution of AIC ONeWD to the population of pulsars, adding the possible ablation of the companion by the pulsar to study the spiders, evaluating the detectability of continuous gravitational waves from the recycled population as it was already done for the normal population by \citet{cbc+21} are other interesting topics to explore. In addition, more constraints could be found about the numerous parameters of the model, while focusing on a restricted number of parameters.

\begin{acknowledgements}
This work has been supported by the French Research Agency grant ANR-20-CE31-0010. We would like to thank Francesca Calore, Joanna Berteaud and Maïca Clavel for insightful discussions. \matteo{D.M. acknowledges the support of the Department of Atomic Energy, Government of India, under project No. 12-R\&D-TFR-5.02-0700. We also thank the anonymous referee for the helpful comments and suggestions, which helped improve the paper.} 
\end{acknowledgements}

\bibliographystyle{aa}
\bibliography{biblio}

\begin{appendix}

\section{Ratio of the pulsar's radiation loss torque and the gravitational wave torque} \label{appendix:AppA}

In the set of differential equations that we use (equation \eqref{eq:diff_eq_model1} to \eqref{eq:diff_eq_model3}), we take into account a torque, $n_{\rm GW}$, caused by GWs. However, this torque depends on $\Omega^5$, therefore compared to the torque due to the pulsar's radiation loss, $n_3$, which depends to $\Omega^3$, it can often be neglected. If we compute the ratio of both torques we obtain
\begin{equation} \label{eq:ratio_torques}
    \frac{n_{\rm GW}}{n_3} = \frac{G I^2 \Omega^2 \epsilon^2 \mu_0}{5 \pi^3 c^2 B^2 R^6}.
\end{equation}
To choose when to consider the torque caused by GWs, we look into when the ratio of equation \eqref{eq:ratio_torques} would be greater than 0.05. If we take $B=7\times10^4$~T and $\epsilon=10^{-7}$, it gives $\Omega_{\rm lim} \sim 1100$~rad.s$^{-1}$, while actually most of the time $\epsilon$ can be lower than this, or the magnetic field a bit higher, resulting in higher $\Omega_{\rm lim}$. Thus, to be very cautious and thorough, we even adopted $\Omega_{\rm lim} = 960$~rad.s$^{-1}$. Therefore in this work, we do not neglect the GWs torque for $\Omega \geq \Omega_{\rm lim}$. 

\section{Influence of the ISM on the radio pulse profile} \label{appendix:AppB}

Dispersion and scattering caused by the ISM lead to a broader radio pulse profile. In order to determine the influence of the ISM through $\tau_{\rm DM}$, the formula of \citet{blr+14} can be used,
\begin{equation} \label{eq:tau_dm_eq}
\tau_{DM}=\frac{e^2 \Delta f_{\rm ch} \, DM}{4\,\pi^2\,\epsilon_0 \, m_e \, c \, f^3} = 8.3\times10^{15} ~s \, \frac{\Delta f_{\rm ch}}{f^3} \, DM
.\end{equation}
Here, $e$ is the electron charge, $m_e$ its mass, $\Delta f_{\rm ch}$ the width of frequency of the instrument channel in Hz, $f$ the observing frequency in Hz, and $DM$ the dispersion measure in pc/cm$^3$. Furthermore, in order to determine the influence of scattering by an heterogeneous and turbulent ISM through $\tau_{\rm scat}$, the empirical fit relationship from \citet{kmn+15} is used
\begin{equation} \label{eq:tau_scat_eq}
\tau_{\rm scat} = 3.6\times 10^{-9} DM^{2.2} (1+1.94\times 10^{-3} DM^2) \ .
\end{equation}
Both $\tau_{\rm DM}$ and $\tau_{\rm scat}$ are given in units of second (s) and depend on the dispersion measure $DM$ (in units of pc/cm$^3$), which is found for each pulsar by running the code of \citet{ymw17} converting the distance of a pulsar into the dispersion measure thanks to their state-of-the art model of the Galactic electron density distribution. See Table~\ref{Table:survey_params} for the parameters used for the two surveys considered, i.e., FAST GPPS and PMPS. 

\section{Anisotropy of the striped wind model} \label{appendix:AppC}

\begin{figure}[h]
\resizebox{\hsize}{!}{\includegraphics{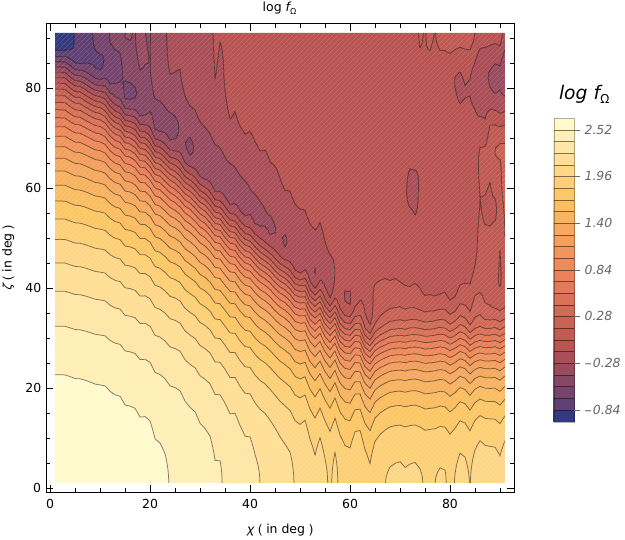}}
\caption{Flux anisotropy factor $f_{\rm \Omega}$ (in log) for the striped wind model versus obliquity $\chi$ and inclination of Earth line of sight $\zeta$.}
\label{fig:fomega}
\end{figure}

The link between the $\gamma$-ray flux and the total $\gamma$-ray luminosity is not a straightforward relation to deduce. The problem resides in the highly anisotropic emission pattern of the pulsar striped wind. This anisotropy is usually embedded in a so-called beaming factor $f_\Omega$ describing the flux pattern depending on the obliquity $\chi$ and line of sight inclination $\zeta$. Its definition is found in \cite{wrw+09,p11,p25} and shown on a log scale in fig.~\ref{fig:fomega} for the force-free dipole magnetosphere model. For high inclinations $\zeta$ it is about $0.2$ whereas for low inclinations $\zeta$, it is very high, well above 10.

\section{SKA survey parameters} \label{appendix:AppD}

Table~\ref{Table:survey_params_SKA} displays the SKA survey parameters in the mid-frequency ranges we used to make the predictions in Sect.~\ref{subsec:detec_prospects}. SKA is described in \citet{cr04,ks15,gaa+17}. SKA-1-Mid represents the estimate of the initially planned SKA operation. 

\begin{table}[h]
\caption{Survey parameters of SKA-1-Mid.} 
\label{Table:survey_params_SKA} 
\centering 
\begin{tabular}{c c} 
\hline\hline 
Survey & SKA-1-Mid  \\
\hline
Sky region & -90° < $b$ < 30°\\
$f$ (GHz) & 1.400\\
$\Delta f_{ch}$ (kHz)& 9\\
$\tau_{\rm samp}$ (s) & $6.4\times10^{-5}$ \\
$G$ (K.Jy$^{-1}$) & 15 \\
$N_p$ & 2 \\
$B$ (MHz) & 300\\
$t$ (s) & 2100\\
$T_{\rm sys}$ & 30\\
$C_{\rm thres}$ & 9\\
\hline 
\end{tabular}
\end{table}

\end{appendix}

\end{document}